\documentclass[preprint,nonatbib]{elsarticle}

\usepackage{lineno,hyperref}
\usepackage{latexsym}
\usepackage{graphicx}
\usepackage{multicol,multirow}
\usepackage{amsmath,amssymb,amsfonts}
\usepackage{mathrsfs}
\usepackage{amsthm}
\usepackage{subcaption}
\usepackage{upgreek}
\usepackage{color}
\usepackage[dvipsnames]{xcolor}

\usepackage{algorithm}
\usepackage{algpseudocode}

\definecolor{Black}{rgb}{0,0,0}
\definecolor{Blue}{rgb}{0,0,1}
\definecolor{Red}{rgb}{1,0,0}
\definecolor{Green}{rgb}{0,1,0}
\definecolor{Cyan}{rgb}{0,0.72,0.92}
\definecolor{Amethyst}{rgb}{0.6,0.4,0.8}
\definecolor{Bronze}{rgb}{0.8,0.5,0.2}
\definecolor{Violet}{rgb}{0.54,0.17,0.89}

\makeatletter
\let\c@author\relax
\makeatother
\usepackage[style=ieee, citestyle=numeric-comp, backend=bibtex]{biblatex}
\begin{document}

\begin{frontmatter}

\title{Physics-informed active learning with simultaneous weak-form latent space dynamics identification}
\date{}

\author[1]{Xiaolong He\corref{corresponding_author}}
\cortext[corresponding_author]{Corresponding author}
\ead{xiaolong.he@ansys.com}
\author[2]{April Tran}
\author[2]{David M. Bortz}
\author[3]{Youngsoo Choi}

\address[1]{ANSYS Inc., Livermore, CA, 94551, USA}
\address[2]{Department of Applied Mathematics, University of Colorado, Boulder, CO, 80309, USA}
\address[3]{Center for Applied Scientific Computing, Lawrence Livermore National Laboratory, Livermore, CA, 94550, USA}





\begin{abstract}
The parametric greedy latent space dynamics identification (gLaSDI) framework has demonstrated promising potential for accurate and efficient modeling of high-dimensional nonlinear physical systems. However, it remains challenging to handle noisy data. To enhance robustness against noise, we incorporate the weak-form estimation of nonlinear dynamics (WENDy) into gLaSDI. 
In the proposed weak-form gLaSDI (WgLaSDI) framework, an autoencoder and WENDy are trained simultaneously to discover intrinsic nonlinear latent-space dynamics of high-dimensional data.
Compared to the standard sparse identification of nonlinear dynamics (SINDy) employed in gLaSDI, WENDy enables variance reduction and robust latent space discovery, therefore leading to more accurate and efficient reduced-order modeling.
Furthermore, the greedy physics-informed active learning in WgLaSDI enables adaptive sampling of optimal training data on the fly for enhanced modeling accuracy.
The effectiveness of the proposed framework is demonstrated by modeling various nonlinear dynamical problems, including viscous and inviscid Burgers' equations, time-dependent radial advection, and the Vlasov equation for plasma physics. 
With data that contains 5-10$\%$ Gaussian white noise, WgLaSDI outperforms gLaSDI by orders of magnitude, achieving 1-7$\%$ relative errors. 
Compared with the high-fidelity models, WgLaSDI achieves 121 to 1,779$\times$ speed-up.
\end{abstract}
\end{frontmatter}


\section{Introduction}\label{sec:introduction}
Physical simulations have played an increasingly significant role in the development of various scientific and engineering domains, including physics, biology, electronics, automotive, aerospace, and digital twin \cite{thijssen2007computational,noble2002rise,vasileska2017computational,bondeson2012computational,muhammad2019simulation,kurec2019advanced,cummings2015applied}.
Reduced-order modeling has been developed to accelerate simulations while maintaining high accuracy to address computational challenges arising from limited computational resources and complex physical phenomena.
One classical method is the projection-based reduced-order model (ROM), which can be categorized into \textit{linear} and \textit{nonlinear} projections. 
\textit{Linear} projection techniques, such as the proper orthogonal decomposition \cite{berkooz1993proper}, the reduced basis method \cite{patera2007reduced}, and the balanced truncation method \cite{safonov1989schur}, have been successfully applied to a wide range of applications, including fluid dynamics \cite{iliescu2014variational,stabile2018finite,copeland2022reduced,cheung2023local,lauzon2022sopt}, fracture mechanics \cite{chen2015model,he2019decomposed}, topology optimization \cite{gogu2015improving,choi2019accelerating}, structural design optimization \cite{mcbane2021component,choi2020gradient}, etc.
However, the linear projection-based ROM falls short when dealing with problems whose Kolmogorov width decays slow, e.g., advection-dominated systems. 
In contrast, \textit{nonlinear} projection techniques based on autoencoders have demonstrated better performance for these systems \cite{kim2022fast,fries2022lasdi,he2023glasdi,diaz2024fast,wlasdi,bonneville2024gplasdi,bonneville2024comprehensive}.

While most projection-based ROMs (pROMs) are \textit{intrusive}, maintaining extrapolation robustness and high accuracy with less training data, they necessitate access to the numerical solver and a detailed understanding of specific implementation.
On the other hand, \textit{non-intrusive} ROMs \cite{kutz2017deep,min2017deep,paganini2018calogan,morton2018deep,kadeethum2021framework,kadeethum2022continuous,kadeethum2022non,kadeethum2022reduced,kim2019deep} are purely data-driven, independent of governing equations of physics and the high-fidelity physical solver.
However, many non-intrusive ROMs lack interpretability, leading to unstable generalization performance.

To address these issues, equation learning algorithms \cite{schmidt2009distilling,brunton2016discovering,peherstorfer2016data} have been integrated into ROMs to identify the underlying equations governing the latent-space dynamics \cite{qian2020lift,benner2020operator,cranmer2020discovering,issan2022predicting}. 
Champion, et al. \cite{champion2019data} applied an autoencoder for dimensionality reduction and the \textit{Sparse Identification of Nonlinear Dynamics} (SINDy) method to identify ordinary differential equations (ODEs) that govern the latent-space dynamics, where the autoencoder and the SINDy model were trained simultaneously to achieve simple latent-space dynamics.
However, the method is not parameterized and generalizable.
Recently, Fries, et al. \cite{fries2022lasdi} proposed a parametric \textit{Latent Space Dynamics Identification} (LaSDI) framework in which an autoencoder was applied for nonlinear projection and a set of SINDy models were employed to identify ODEs governing the latent-space dynamics of the training data.
During inference, the ODE coefficients of the testing case are obtained by interpolating those of training data, allowing for estimating the latent space dynamics of any new parameter in the parameter space.
Recent developments \cite{bonneville2024comprehensive} have extended LaSDI to achieve simultaneous training and active learning based on the residual error (gLaSDI \cite{he2023glasdi}), active learning based on Gaussian processes \cite{bonneville2024gplasdi}, and to follow the thermodynamics principles \cite{park2024tlasdi}.



The LaSDI class of algorithms relies on SINDy for learning the ODEs governing the latent space dynamics.\footnote{excluding tLaSDI \cite{park2024tlasdi}} SINDy \cite{brunton2016discovering}, however, is widely acknowledged to be sensitive to noise in the data, due to the pointwise derivative approximation.
To address this sensitivity, weak form methods such as weak SINDy (WSINDy) for equation learning (of ODEs \cite{wsindy1,MessengerWheelerLiuEtAl2022JRSocInterface,MessengerBurbyBortz2024SciRep}, PDEs \cite{wsindy2}, SDEs \cite{MessengerBortz2022PhysicaD}, and hybrid systems \cite{MessengerDwyerDukic2024arXiv240520591}) and WENDy for parameter estimation \cite{wendy} have been developed with strong theoretical convergence performance \cite{MessengerBortz2024IMAJNumerAnal}.  When the data is noisy, the latent variables constructed in LaSDI (autoencoder or POD) could still retain that noise. Accordingly, Tran et al., \cite{wlasdi} proposed the WLaSDI method, extending LaSDI to use WENDy for system identification of the latent dynamics.  This results in a ROM creation method that is robust to large amounts of noise while maintaining the ability to create an accurate ROM that can be simulated hundreds of times faster than the full-order model.

In WLaSDI, however, a two-step sequential procedure was adopted, first training the autoencoder and then using WENDy to estimate parameters in the ODE models for the latent dynamics. The lack of interaction between them could result in complex latent space representation discovered by the autoencoder, which could further pose challenges to the subsequent equation learning by WENDy and thus affect the model performances. 

\begin{figure}[htp]
    \centering
    \includegraphics[width=0.8\linewidth]{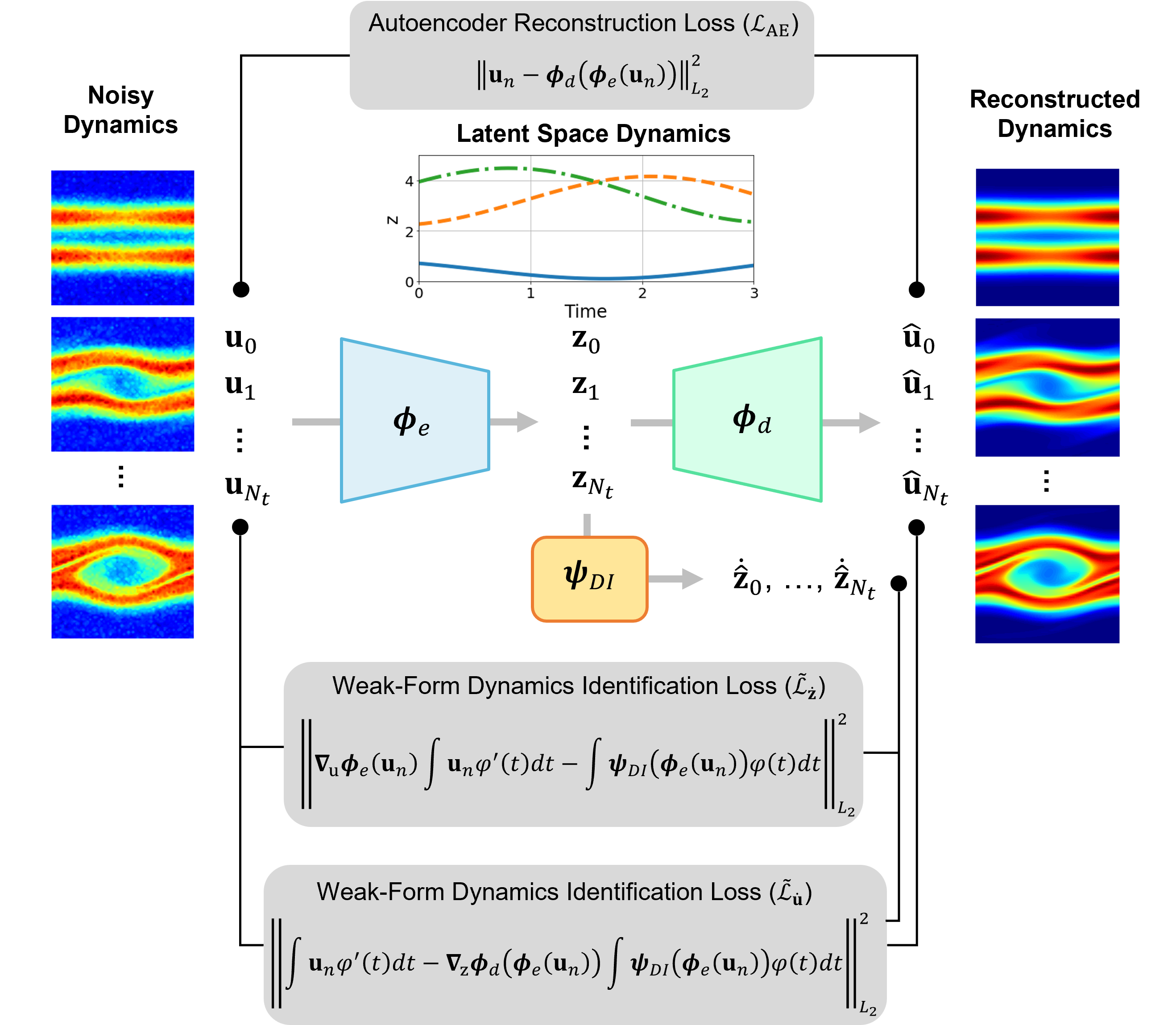}
    \caption{Schematics of the interactive training of the autoencoder and WENDy in proposed WgLaSDI framework to counteract noisy data. Note that the introduction of the weak form reduces the data's variance and avoids the need for the noisy data's time derivative.}
    \label{fig.wglasdi}
\end{figure}

To overcome this limitation, we propose a weak-form greedy latent space dynamics identification (WgLaSDI) framework (Fig. \ref{fig.wglasdi}).
WgLaSDI is directly built upon gLaSDI, which includes simultaneous autoencoder training and dynamics identification, $k$-NN convex interpolation of latent space dynamics for inference, and greedy physics-informed active learning for selecting optimal training samples.
In WgLaSDI, WENDy is employed for robust latent space dynamics identification and trained simultaneously with the autoencoder to discover intrinsic latent representations.
Each training sample has its own set of WENDy coefficients. 
During inference, the WENDy coefficient of the testing case is estimated by interpolating the WENDy coefficients of $k$ nearest neighbor ($k$-NN) training samples.
The greedy active learning based on the residual error maximizes the diversity of training samples in the parameter space and minimizes prediction errors.
The effectiveness and enhanced performance of the proposed WgLaSDI framework in the presence of noisy data is demonstrated by modeling various nonlinear dynamical problems with a comparison with the gLaSDI and WLaSDI frameworks \cite{he2023glasdi,wlasdi}.

The remainder of this paper is structured as follows. 
The governing equations of physical systems are introduced in Section \ref{sec:govern_eqn}.
In Section \ref{sec:wglasdi}, we introduce the building blocks of gLaSDI and the proposed WgLaSDI framework, including autoencoders, strong-form and weak-form latent space dynamics identification, $k$-NN interpolation, and greedy physics-informed active learning based on a residual error indicator. 
We then demonstrate the effectiveness and capability of the proposed WgLaSDI framework in Section \ref{sec:result} by modeling various nonlinear dynamical problems, including Burgers' equations, time-dependent radial advection, and a two-stream plasma instability problem. 
The effects of the weak form, simultaneous training, and greedy physics-informed active learning are discussed with a comparison with gLaSDI and WLaSDI.
Concluding remarks and discussions are summarized in Section \ref{sec:conclusion}.

\section{Governing Equations of Physical Systems}\label{sec:govern_eqn}
We consider physical systems characterized by a system of partial differential equations (PDEs),
\begin{equation}\label{eq.govern_pde}
    \begin{cases}
    \displaystyle\frac{\partial\Tilde{\mathbf{u}}(t,\mathbf{x}; \boldsymbol{\mu})}{\partial t} = \mathbf{f}(\Tilde{\mathbf{u}},t,\mathbf{x}; \boldsymbol{\mu}), \quad t \in [0,T], \quad \mathbf{x} \in \Omega, \\
    \Tilde{\mathbf{u}}(t=0,\mathbf{x}; \boldsymbol{\mu}) = \Tilde{\mathbf{u}}_0(\mathbf{x}; \boldsymbol{\mu}), \quad \boldsymbol{\mu} \in \mathcal{D}, \\
    \end{cases}
\end{equation}
where $\Tilde{\mathbf{u}}(t,\mathbf{x}; \boldsymbol{\mu})$ represents the solution field (either scalar or vector) over the spatiotemporal domain $\Omega \times [0,T]$ of the physical problem; 
$T \in \mathbb{R}_{+}$ is the final time of the physical problem; 
$\Omega$ is a spatial domain;
$\mathbf{f}$ denotes a differential operator that could include linear or nonlinear spatial derivatives of $\Tilde{\mathbf{u}}$ as well as source terms; 
$\Tilde{\mathbf{u}}_0$ is the initial state of $\Tilde{\mathbf{u}}$, parameterized by $\boldsymbol{\mu} \in \mathcal{D} \subseteq \mathbb{R}^{n_{\mu}}$ that can have any dimension. The parameter space is denoted by $\mathcal{D}$.
We consider a two-dimensional parameter space in this study, i.e., $n_{\mu}=2$, although our method can be applied to problems with any positive integer value of $n_{\mu}$.

The numerical solution to Eq. (\ref{eq.govern_pde}) can be obtained by numerical methods, such as the finite element method. We can obtain a residual function from the discrete system of equations. For example, let $\mathbf{u}(t; \boldsymbol{\mu}) \in \mathbb{R}^{N_u}$ represent the semi-discretized full-order model (FOM) solution at time $t$ and $N_u$ is the degrees of freedom. With the implicit backward Euler time integrator, the approximate solution to Eq. (\ref{eq.govern_pde}) can be obtained by solving the following nonlinear discrete system of equations
\begin{equation}\label{eq.backward_euler}
    \mathbf{u}_n = \mathbf{u}_{n-1} + \Delta t \mathbf{f}_n,
\end{equation}
where $\mathbf{u}_n := \mathbf{u}(t_n; \boldsymbol{\mu}) \in \mathbb{R}^{N_u}$ is the FOM solution at the $n$-th time step and 
$\mathbf{f}_n := \mathbf{f}(\mathbf{u}_n, t_n; \boldsymbol{\mu})$; 
$\Delta t \in \mathbb{R}_{+}$ denotes the time step size and $t_n = t_{n-1} + \Delta t$ for $n \in \mathbb{N}(N_t)$ where $t_0 = 0$ and $N_t \in \mathbb{N}$ is the number of time steps. 
The residual function of Eq. (\ref{eq.backward_euler}) is expressed as
\begin{equation}\label{eq.residual}
    \mathbf{r}(\mathbf{u}_n; \mathbf{u}_{n-1}, \boldsymbol{\mu}) = \mathbf{u}_n - \mathbf{u}_{n-1} - \Delta t \mathbf{f}_n.
\end{equation}

Given a parameter $\boldsymbol{\mu}^{(i)} \in \mathcal{D}$, $i \in \mathbb{N}(N_{\mu})$, we denote $\mathbf{u}_n^{(i)}$ as the solution at the $n$-th time step of the physical system defined in Eq. (\ref{eq.govern_pde}). 
The solutions at all time steps are arranged in a snapshot matrix denoted as $\mathbf{U}^{(i)} = [\mathbf{u}_0^{(i)}, ..., \mathbf{u}_{N_t}^{(i)}]^\top \in \mathbb{R}^{(N_t + 1) \times N_u}$. Compiling snapshot matrices corresponding to all parameters gives a complete FOM dataset $\mathbf{U} \in \mathbb{R}^{N_{\mu} \times (N_t + 1) \times N_u}$.

In practical applications with complex physical phenomena, it is computationally expensive to solve the system of equation (Eq. (\ref{eq.govern_pde})), especially when the degrees of freedom ($N_u$) is large and the computational domain ($\Omega$) is geometrically complex. In this work, we aim to develop an accurate, efficient, and robust reduced-order modeling framework based on latent space dynamics identification, which will be discussed in detail in the following sections.

\section{Weak-Form Greedy Latent Space Dynamics Identification (WgLaSDI)}\label{sec:wglasdi}

In this section, we introduce the building blocks of the proposed WgLaSDI framework, including autoencoders, strong-form and weak-form latent space dynamics identification, $k$-NN interpolation, greedy physics-informed active learning based on a residual error indicator, and generalization. 

\subsection{Autoencoders for Nonlinear Dimensionality Reduction}\label{sec:autoencoders}
An autoencoder \cite{demers1993non,hinton2006reducing} is a type of neural network designed for dimensionality reduction or representation (manifold) learning. It contains an encoder function $\boldsymbol{\phi}_e(\cdot;\boldsymbol{\theta}_{\text{enc}}) : \mathbb{R}^{N_u} \rightarrow \mathbb{R}^{N_z}$ and a decoder function $\boldsymbol{\phi}_d(\cdot;\boldsymbol{\theta}_{\text{dec}}): \mathbb{R}^{N_z} \rightarrow \mathbb{R}^{N_u}$, such that
\begin{subequations}\label{eq.autoencoder}
    \begin{align}
        & \mathbf{z}_n^{(i)} = \boldsymbol{\phi}_e(\mathbf{u}_n^{(i)};\boldsymbol{\theta}_{\text{enc}}), \\
        & \hat{\mathbf{u}}_n^{(i)} = \boldsymbol{\phi}_d(\mathbf{z}_n^{(i)};\boldsymbol{\theta}_{\text{dec}}),
    \end{align}
\end{subequations}
where $\boldsymbol{\theta}_{\text{enc}}$ and $\boldsymbol{\theta}_{\text{dec}}$ are network parameters of the encoder and the decoder, respectively.
The encoder transforms the high-dimensional input data $\mathbf{u}_n^{(i)} \in \mathbb{R}^{N_u}$ to a low-dimensional latent representation $\mathbf{z}_n^{(i)} \in \mathbb{R}^{N_z}$, where $N_z$ is the latent dimension. Typically, $N_z \ll N_u$ is chosen to achieve dimensionality reduction. 
The decoder transforms the compressed data $\mathbf{z}_n^{(i)}$ to a reconstructed version of $\mathbf{u}_n^{(i)}$, denoted as $\hat{\mathbf{u}}_n^{(i)} \in \mathbb{R}^{N_u}$.
 
Given a FOM dataset $\mathbf{U}$, we can obtain the latent and reconstructed representations through the encoder and the decoder, denoted as $\mathbf{Z} \in \mathbb{R}^{ N_{\mu} \times (N_t + 1) \times N_z}$ and  $\hat{\mathbf{U}} \in \mathbb{R}^{N_{\mu} \times (N_t + 1) \times N_u}$, respectively. 
The optimal network parameters of the autoencoder ($\boldsymbol{\theta}_{\text{enc}}$ and $\boldsymbol{\theta}_{\text{dec}}$ from Eq. (\ref{eq.autoencoder})) can be obtained by minimizing the loss function:
\begin{equation}\label{eq.autoencoder_loss}
\begin{aligned}
    \mathcal{L}_{\text{AE}}(\boldsymbol{\theta}_{\text{enc}}, \boldsymbol{\theta}_{\text{dec}}) 
    & := \big|\big| \mathbf{U} - \hat{\mathbf{U}} \big|\big|_{L_2}^2 \\
    & = \frac{1}{N_{\mu}}\sum_{i=1}^{N_{\mu}} 
    \bigg( 
    \frac{1}{N_{t}+1}\sum_{n=0}^{N_t} \big|\big| \mathbf{u}_n^{(i)} - 
    \boldsymbol{\phi}_d(\boldsymbol{\phi}_e(\mathbf{u}_n^{(i)};\boldsymbol{\theta}_{\text{enc}});\boldsymbol{\theta}_{\text{dec}}) \big|\big|_{L_2}^2 
    \bigg).
\end{aligned}
\end{equation}

\subsection{Identification of Latent Space Dynamics}\label{sec:dyn_ident}
\subsubsection{Strong Form}\label{sec:strong_form}
The low-dimensional latent variable ($\mathbf{z}$) discovered by the autoencoder inherits the time dependency from the high-dimensional FOM variable ($\mathbf{u}$). 
The dynamics of the latent variable in the latent space can be described by an equation of the following form,
\begin{equation}\label{eq.dynamics_identification}
    \frac{d\mathbf{z}(t; \boldsymbol{\mu})}{dt} = \boldsymbol{\psi}_{DI}(\mathbf{z},t; \boldsymbol{\mu}), \quad t \in [0,T],
\end{equation}
where $\mathbf{z}(t; \boldsymbol{\mu}) = \boldsymbol{\phi}_e(\mathbf{u}(t; \boldsymbol{\mu}))$ and $\boldsymbol{\psi}_{DI}$ represents a \textit{Dynamics Identification} (DI) function governing the latent space dynamics. 
$\boldsymbol{\psi}_{DI}$ can be approximated by a system of ODEs and identified by the sparse identification of nonlinear dynamics (SINDy) method \cite{brunton2016discovering}. 
\begin{equation}\label{eq.dynamics_identification_discrete}
     \dot{\mathbf{z}}_n^{(i)} \approx \dot{\hat{\mathbf{z}}}_n^{(i)} = \boldsymbol{\Theta}(\mathbf{z}_n^{(i)}) \boldsymbol{\Xi}^{(i)},
\end{equation}
where $\boldsymbol{\Theta}(\mathbf{z}_n^{(i)}) \in \mathbb{R}^{N_l}$ is a user-defined library that contains ${N_l}$ linear and nonlinear terms \cite{brunton2016discovering,fries2022lasdi,he2023glasdi};
$\boldsymbol{\Xi}^{(i)} \in \mathbb{R}^{N_l \times N_z}$ is a coefficient matrix associated with the parameter $\boldsymbol{\mu}^{(i)} \in \mathcal{D}$. 
The time derivative $\dot{\mathbf{z}}_n^{(i)}$ can be obtained by using the chain rule:
\begin{equation}\label{eq.zdot}
    \dot{\mathbf{z}}_n^{(i)} = 
    \frac{\partial \mathbf{z}_n^{(i)}}{\partial \mathbf{u}_n^{(i)}} \dot{\mathbf{u}}_n^{(i)} = 
    \frac{\partial \boldsymbol{\phi}_e(\mathbf{u}_n^{(i)};\boldsymbol{\theta}_{\text{enc}})}{\partial \mathbf{u}_n^{(i)}} \dot{\mathbf{u}}_n^{(i)}.
\end{equation}
The coefficients of all parameters $\boldsymbol{\Xi} = \big[ \boldsymbol{\Xi}^{(1)}, ... , \boldsymbol{\Xi}^{(N_{\mu})} \big] \in \mathbb{R}^{N_{\mu} \times N_l \times N_z}$ can be obtained by minimizing the following function:
\begin{equation}\label{eq.loss_zdot}
\begin{aligned}
    \mathcal{L}_{\dot{\mathbf{z}}}(\boldsymbol{\theta}_{\text{enc}}, \boldsymbol{\Xi}) 
    & := \big|\big| \dot{\mathbf{Z}} - \dot{\hat{\mathbf{Z}}} \big|\big|_{L_2}^2 \\
    & =  \frac{1}{N_{\mu}}\sum_{i=1}^{N_{\mu}} 
    \bigg( 
    \frac{1}{N_{t}+1}\sum_{n=0}^{N_t} 
    \bigg|\bigg| 
    \frac{\partial \boldsymbol{\phi}_e(\mathbf{u}_n^{(i)};\boldsymbol{\theta}_{\text{enc}})}{\partial \mathbf{u}_n^{(i)}} \dot{\mathbf{u}}_n^{(i)} - 
    \boldsymbol{\Theta}(\mathbf{z}_n^{(i)}) \boldsymbol{\Xi}^{(i)} 
    \bigg|\bigg|_{L_2}^2 
    \bigg) 
\end{aligned}
\end{equation}
This loss function $\mathcal{L}_{\dot{\mathbf{z}}}$ provides constraints to the encoder-predicted latent space dynamics ($\dot{\mathbf{Z}}$) such that it is consistent with that from the DI model ($\dot{\hat{\mathbf{Z}}}$), which is critical for discovery of simple and smooth latent dynamics and therefore model performance \cite{he2023glasdi}.

To further enhance the accuracy of the physical dynamics predicted by the decoder, the following loss function is constructed to ensure the consistency between the predicted dynamics gradients ($\dot{\hat{\mathbf{u}}}_n^{(i)}$) and the gradients of the solution data ($\dot{\mathbf{u}}_n^{(i)}$):
\begin{equation}\label{eq.loss_udot}
\begin{aligned}
    \mathcal{L}_{\dot{\mathbf{u}}}(\boldsymbol{\theta}_{\text{enc}}, \boldsymbol{\theta}_{\text{dec}}, \boldsymbol{\Xi}) 
    & := || \dot{\mathbf{U}} - \dot{\hat{\mathbf{U}}} ||_{L_2}^2 \\
    & = \frac{1}{N_{\mu}}\sum_{i=1}^{N_{\mu}} 
    \bigg( 
    \frac{1}{N_{t}+1}\sum_{n=0}^{N_t} 
    \big|\big| 
    \dot{\mathbf{u}}_n^{(i)} - \dot{\hat{\mathbf{u}}}_n^{(i)}
    \big|\big|_{L_2}^2 
    \bigg),
\end{aligned}
\end{equation}
where $\dot{\hat{\mathbf{u}}}_n^{(i)}$ can be obtained by applying the chain rule: 
\begin{equation}\label{eq.udot}
    \dot{\hat{\mathbf{u}}}_n^{(i)} 
    = \frac{\partial \hat{\mathbf{u}}_n^{(i)}}{\partial \mathbf{z}_n^{(i)}}  \dot{\mathbf{z}}_n^{(i)}
    = \frac{\partial \boldsymbol{\phi}_d \big( \boldsymbol{\phi}_e(\mathbf{u}_n^{(i)};\boldsymbol{\theta}_{\text{enc}}) ;\boldsymbol{\theta}_{\text{dec}}\big)}{\partial \mathbf{z}_n^{(i)}}  \boldsymbol{\Theta}(\mathbf{z}_n^{(i)}) \boldsymbol{\Xi}^{(i)}.
\end{equation}
Therefore, the loss function of gLaSDI is defined as
\begin{equation}\label{eq.total_loss}
    \mathcal{L}(\boldsymbol{\theta}_{\text{enc}}, \boldsymbol{\theta}_{\text{dec}}, \boldsymbol{\Xi}) = \mathcal{L}_{\text{AE}}(\boldsymbol{\theta}_{\text{enc}}, \boldsymbol{\theta}_{\text{dec}}) + 
    \beta_1 \mathcal{L}_{\dot{\mathbf{z}}}(\boldsymbol{\theta}_{\text{enc}}, \boldsymbol{\Xi})  + 
    \beta_2 \mathcal{L}_{\dot{\mathbf{u}}}(\boldsymbol{\theta}_{\text{enc}}, \boldsymbol{\theta}_{\text{dec}}, \boldsymbol{\Xi}),
\end{equation}
where $\beta_1$ and $\beta_2$ denote the regularization parameters to balance the scale and contributions from the loss terms. 

\subsubsection{Weak Form}\label{sec:weak_form}
The latent space dynamics identification described in Section \ref{sec:strong_form} is based on the strong-form ODE. If the high-dimensional input data contains noise and errors, the latent representation could be nonlinear and oscillatory, which poses challenges to the strong-form dynamics identification. To enhance the robustness against noise, we leverage the variance-reduction nature of the weak form and integrate it into gLaSDI.

We begin by multiplying Eqs.~\eqref{eq.dynamics_identification_discrete} and \eqref{eq.zdot} with a continuous and compactly supported test function $\varphi(t)\in \mathcal{C}_c(\mathbb{R},\mathbb{R})$ and integrating them over a time domain:
\begin{equation}\label{eq.weak_dynamics_identification}
    \int_{t_a}^{t_b} \dot{\hat{\mathbf{z}}}_n^{(i)} \varphi(t)dt  =  
    \int_{t_a}^{t_b} \boldsymbol{\Theta}(\mathbf{z}_n^{(i)}) \mathbf{\Xi}^{(i)} \varphi(t) dt,
\end{equation}

\begin{equation}\label{eq.weak_zdot1}
    \int_{t_a}^{t_b} \dot{\mathbf{z}}_n^{(i)} \varphi(t)dt = 
    \int_{t_a}^{t_b} \frac{\partial \mathbf{z}_n^{(i)}}{\partial \mathbf{u}_n^{(i)}} \dot{\mathbf{u}}_n^{(i)} \varphi(t)dt.
\end{equation}
When a piece-wise linear activation function, such as ReLU, is employed in the encoder network, $\partial \mathbf{z}_n^{(i)}/\partial \mathbf{u}_n^{(i)}$ is piece-wise constant, which is independent of time and can be taken out of the time integral in Eq. \eqref{eq.weak_zdot1}:
\begin{equation}\label{eq.weak_zdot}
    \int_{t_a}^{t_b} \dot{\mathbf{z}}_n^{(i)} \varphi(t)dt =
    \frac{\partial \mathbf{z}_n^{(i)}}{\partial \mathbf{u}_n^{(i)}}
    \int_{t_a}^{t_b} \dot{\mathbf{u}}_n^{(i)} \varphi(t)dt.
\end{equation}
Due to the compact support of $\varphi$, we know that $\varphi(t_b) =  \varphi(t_a) = 0$, and applying integration by parts yields:
\begin{equation}\label{eq.weak_zdot2}
    \int_{t_a}^{t_b} \dot{\mathbf{z}}_n^{(i)} \varphi(t)dt 
    =
    - \frac{\partial \mathbf{z}_n^{(i)}}{\partial \mathbf{u}_n^{(i)}}
    \int_{t_a}^{t_b} \mathbf{u}_n^{(i)} \varphi'(t)dt.
\end{equation}
We apply the trapezoidal rule to discretize the time integrals. 
Thus, the weak-form transformation allows $\int_{t_a}^{t_b} \dot{\mathbf{u}}_n^{(i)} \varphi(t)dt$ to be computed without using the time derivative of the data ($\dot{\mathbf{u}}_n^{(i)}$), which is challenging to obtain in the presence of noise. 
It contributes to the enhanced accuracy and stability of latent space dynamics identification, as will be demonstrated in Section \ref{sec:result}.
We note that alternatively, $\int_{t_a}^{t_b} \dot{\mathbf{z}}_n^{(i)} \varphi(t)dt$ can be computed by directly applying integration by parts:
\begin{equation}\label{eq.weak_zdot3}
    \int_{t_a}^{t_b} \dot{\mathbf{z}}_n^{(i)} \varphi(t)dt
    = 
    - \int_{t_a}^{t_b} \mathbf{z}_n^{(i)} \varphi'(t)dt.
\end{equation}

The weak-form loss function $\tilde{\mathcal{L}}_{\dot{\mathbf{z}}}$ corresponding to the strong-form counterpart $\mathcal{L}_{\dot{\mathbf{z}}}$ that enforces $\dot{\mathbf{z}}_n^{(i)} = \dot{\hat{\mathbf{z}}}_n^{(i)}$, as defined in Eq. \eqref{eq.loss_zdot}, becomes
\begin{equation}\label{eq.weak_loss_zdot}
    \tilde{\mathcal{L}}_{\dot{\mathbf{z}}}(\boldsymbol{\theta}_{\text{enc}}, \boldsymbol{\Xi}) 
    := \frac{1}{N_{\mu}}\sum_{i=1}^{N_{\mu}} 
    \bigg( 
    \frac{1}{N_{t}+1}\sum_{n=0}^{N_t} 
    \bigg|\bigg| 
    \int_{t_a}^{t_b} \dot{\mathbf{z}}_n^{(i)} \varphi(t)dt 
    - \int_{t_a}^{t_b} \dot{\hat{\mathbf{z}}}_n^{(i)} \varphi(t)dt
    \bigg|\bigg| _{L_2}^2
    \bigg),
\end{equation}
where $\int_{t_a}^{t_b} \dot{\hat{\mathbf{z}}}_n^{(i)} \varphi(t)dt$ is obtained by Eq. \eqref{eq.weak_dynamics_identification} and $\int_{t_a}^{t_b} \dot{\mathbf{z}}_n^{(i)} \varphi(t)dt$ can be computed by either Eq. \eqref{eq.weak_zdot2} or \eqref{eq.weak_zdot3}. 
We will compare the effects of these two approaches on the model performance in Section \ref{sec:result}.

Similarly, we can derive the weak-form loss function $\tilde{\mathcal{L}}_{\dot{\mathbf{u}}}$ corresponding to the strong-form counterpart $\mathcal{L}_{\dot{\mathbf{u}}}$ defined in Eq. \eqref{eq.loss_udot} that enforces
\begin{equation}\label{eq.udot_eqn}
    \dot{\mathbf{u}}_n^{(i)} = \dot{\hat{\mathbf{u}}}_n^{(i)}.
\end{equation}
Multiplying Eq. \eqref{eq.udot_eqn} with the test function $\varphi(t)$ and integrating it over a time domain gives:
\begin{equation}\label{eq.weak_udot_eqn}
    \int_{t_a}^{t_b} \dot{\mathbf{u}}_n^{(i)} \varphi(t) dt
    = 
    \int_{t_a}^{t_b} \dot{\hat{\mathbf{u}}}_n^{(i)} \varphi(t) dt.
\end{equation}
The right-hand side can be obtained from multiplying Eq. \eqref{eq.udot} with the test function $\varphi(t)$ and integrating it over a time domain:
\begin{equation}\label{eq.weak_uhatdot}
\begin{aligned}
    \int_{t_a}^{t_b} \dot{\hat{\mathbf{u}}}_n^{(i)} \varphi(t)dt 
    & = 
    \int_{t_a}^{t_b} \frac{\partial \hat{\mathbf{u}}_n^{(i)}}{\partial \mathbf{z}_n^{(i)}} \boldsymbol{\Theta}(\mathbf{z}_n^{(i)}) \boldsymbol{\Xi}^{(i)} \varphi(t)dt \\ 
    & = 
    \frac{\partial \hat{\mathbf{u}}_n^{(i)}}{\partial \mathbf{z}_n^{(i)}} 
    \int_{t_a}^{t_b} \boldsymbol{\Theta}(\mathbf{z}_n^{(i)}) \boldsymbol{\Xi}^{(i)} \varphi(t)dt,
\end{aligned}
\end{equation}
where $\partial \hat{\mathbf{u}}_n^{(i)}/\partial \mathbf{z}_n^{(i)}$ can be taken out of the time integral when a piece-wise linear activation function is employed in the decoder network.
Alternatively, $\int_{t_a}^{t_b} \dot{\hat{\mathbf{u}}}_n^{(i)} \varphi(t)dt $ can be obtained by directly applying integration by parts:
\begin{equation}\label{eq.weak_uhatdot2}
    \int_{t_a}^{t_b} \dot{\hat{\mathbf{u}}}_n^{(i)} \varphi(t)dt
    = 
    - \int_{t_a}^{t_b} \hat{\mathbf{u}}_n^{(i)} \varphi'(t)dt.
\end{equation}

The weak-form loss function $\tilde{\mathcal{L}}_{\dot{\mathbf{u}}}$ is defined as:
\begin{equation}\label{eq.weak_loss_udot}
    \tilde{\mathcal{L}}_{\dot{\mathbf{u}}}(\boldsymbol{\theta}_{\text{enc}}, \boldsymbol{\theta}_{\text{dec}}, \boldsymbol{\Xi}) 
    := \frac{1}{N_{\mu}}\sum_{i=1}^{N_{\mu}} 
    \bigg( 
    \frac{1}{N_{t}+1}\sum_{n=0}^{N_t} 
    \bigg|\bigg| 
    \int_{t_a}^{t_b} \dot{\mathbf{u}}_n^{(i)} \varphi(t) dt
    -
    \int_{t_a}^{t_b} \dot{\hat{\mathbf{u}}}_n^{(i)} \varphi(t) dt
    \bigg|\bigg| _{L_2}^2
    \bigg),
\end{equation}
where $\int_{t_a}^{t_b} \dot{\mathbf{u}}_n^{(i)} \varphi(t) dt =     - \int_{t_a}^{t_b} \mathbf{u}_n^{(i)} \varphi'(t)dt$ and $\int_{t_a}^{t_b} \dot{\hat{\mathbf{u}}}_n^{(i)} \varphi(t) dt$ can be obtained by either Eq. \eqref{eq.weak_uhatdot} or \eqref{eq.weak_uhatdot2}.
We will compare the effects of these two approaches on the model performance in Section \ref{sec:result}.

The loss function of weak-form gLaSDI is defined as
\begin{equation}\label{eq.weak_total_loss}
\begin{aligned}
    \mathcal{L}(\boldsymbol{\theta}_{\text{enc}}, \boldsymbol{\theta}_{\text{dec}}, \boldsymbol{\Xi}) 
    & = \mathcal{L}_{\text{AE}}(\boldsymbol{\theta}_{\text{enc}}, \boldsymbol{\theta}_{\text{dec}}) + 
    \beta_1 \tilde{\mathcal{L}}_{\dot{\mathbf{z}}}(\boldsymbol{\theta}_{\text{enc}}, \boldsymbol{\Xi}) \\
    & + 
    \beta_2 \tilde{\mathcal{L}}_{\dot{\mathbf{u}}}(\boldsymbol{\theta}_{\text{enc}}, \boldsymbol{\theta}_{\text{dec}}, \boldsymbol{\Xi}) +
    \beta_3 ||\boldsymbol{\Xi}||_{L_2}^2,
\end{aligned}
\end{equation}
where the last term is a sparsity constraint of the ODE coefficients to achieve better-conditioned ODEs for the latent space. 
Fig. \ref{fig.wglasdi} depicts the interactive training of the autoencoder and WENDy in the proposed WgLaSDI framework to counteract noisy data.
In this study, piece-wise polynomials are employed for the test function.
For further details about the selection of test functions, please refer to \cite{wendy,wsindy2,wsindy1}.

\subsection{Interpolation of Latent Space Dynamics}\label{sec:interpolation}
Each training parameter point in the parameter space is attached to a local DI model with its own ODE coefficients. The DI model for a testing parameter point can be obtained by interpolating the local DI model of training parameters. In gLaSDI, a $k$-NN convexity-preserving partition-of-unity interpolation scheme is employed.

Given a testing parameter $\boldsymbol{\mu}^{(*)} \in \mathcal{D}$, the ODE coefficient matrix is obtained by a convex interpolation of coefficient matrices of its $k$-nearest neighbors (existing training parameter points), expressed as
\begin{equation}\label{eq.convex_interp}
    \boldsymbol{\Xi}^{(*)} = \sum_{i \in \mathcal{N}_k(\boldsymbol{\mu}^{(*)})} \Psi^{(i)}(\boldsymbol{\mu}^{(*)}) \boldsymbol{\Xi}^{(i)},
\end{equation}
where $\mathcal{N}_k(\boldsymbol{\mu}^{(*)})$ is a set of indices of the $k$-nearest neighbors (training parameter points) of $\boldsymbol{\mu}^{(*)}$. The selection of the $k$-nearest neighbors is based on the Mahalanobis distance between the testing parameter and the training parameters, $||\boldsymbol{\mu}-\boldsymbol{\mu}^{(i)}||_M$. The interpolation functions are defined as
\begin{equation}\label{eq.shep_shape}
    \Psi^{(i)}(\boldsymbol{\mu}^{(*)}) = \frac{||\boldsymbol{\mu}^{(*)}-\boldsymbol{\mu}^{(i)}||^{-2}_M}{\sum_{j \in \mathcal{N}_k(\boldsymbol{\mu}^{(*)})} ||\boldsymbol{\mu}^{(*)}-\boldsymbol{\mu}^{(j)}||^{-2}_M},
\end{equation}
which satisfy a partition of unity, $\sum_{i \in \mathcal{N}_k(\boldsymbol{\mu})} \Psi^{(i)}(\boldsymbol{\mu}^{(*)})=1$, for transformation objectivity and convexity preservation. For more details and discussion, please refer to \cite{he2023glasdi}.

\subsection{Greedy Physics-Informed Active Learning}\label{sec:active_learning}
To maximally explore the parameter space and achieve optimal model performance, gLaSDI employs a greedy physics-informed active learning strategy to sample training data on the fly. 
Given a testing parameter, the ROM prediction error is evaluated by an error indicator based on the residual of the governing equation:
\begin{equation}\label{eq.residual_norm}
    e_{res} \big( \hat{\mathbf{U}}(\boldsymbol{\mu}^{{(*)}}) \big) 
    = \frac{1}{N_{ts}} \sum_{n=1}^{N_{ts}}||\mathbf{r}(\hat{\mathbf{u}}_n; \hat{\mathbf{u}}_{n-1}, \boldsymbol{\mu}^{(*)})||_{L_2},
\end{equation}
where the residual function $\mathbf{r}(\hat{\mathbf{u}}_n; \hat{\mathbf{u}}_{n-1}, \boldsymbol{\mu})$ is defined in Eq. (\ref{eq.residual}) and $N_{ts}$ denotes the number of time steps used for the evaluation. Typically, $N_{ts} \ll N_t$ is adopted to enhance evaluation efficiency. 

The gLaSDI training is initiated with limited training samples (parameter points). In the examples of this study, the initial training parameter points are located at the corners of the parameter space. Greedy sampling is performed at every $N_{up}$ epochs, at which the current gLaSDI model is evaluated by a finite number of testing parameters based on the error indicator (Eq. \eqref{eq.residual_norm}). The testing parameter associated with the largest error is added to the list of training parameter points. The corresponding FOM data is added to the training dataset before the training resumes. 
This physics-informed active learning process continues until a prescribed number of training samples is reached or the model prediction accuracy is satisfied. Please refer to \cite{he2023glasdi} for more details about the greedy physics-informed active learning algorithms and procedures.

\subsection{Generalization}\label{sec:generalization}
Given a testing parameter $\boldsymbol{\mu}^{{(*)}}$, we first estimate its DI coefficient $\boldsymbol{\Xi}^{(*)}$ by interpolating the DI coefficients of its $k$ nearest neighboring training parameters using Eq. \eqref{eq.convex_interp}) as introduced in Section \ref{sec:interpolation}. After that, we compute the initial condition of the latent variables $\mathbf{z}_0^{(*)} = \boldsymbol{\phi}_e(\mathbf{u}_0^{(*)};\boldsymbol{\theta}_{\text{enc}})$ and predict the latent space dynamics ($\hat{\mathbf{Z}}^{(*)}$) by solving the set of ODE corresponding to $\boldsymbol{\mu}^{{(*)}}$ (Eq. \eqref{eq.dynamics_identification_discrete}) using a numerical integrator. 
Finally, the prediction of the full-order dynamics can be obtained by the decoder, $\hat{\mathbf{U}}^{(*)} = \boldsymbol{\phi}_d(\hat{\mathbf{Z}}^{(*)};\boldsymbol{\theta}_{\text{dec}})$.
\section{Numerical results}\label{sec:result}
The performance of WgLaSDI is demonstrated by solving four numerical problems: a one-dimensional (1D) inviscid Burgers' equation, a two-dimensional (2D) viscous Burgers' equation, time-dependent radial advection, and a two-stream plasma instability problem. In each of the numerical examples, WgLaSDI's performance is compared with that of gLaSDI \cite{he2023glasdi} and WLaSDI \cite{wlasdi} using the maximum relative error defined as:
\begin{equation}\label{eq.max_relative_error}
    e_{max} \big( \mathbf{U}^{(*)}, \hat{\mathbf{U}}^{(*)} \big) 
    = \underset{n}{\text{max}} \Bigg( \frac{\big|\big|\mathbf{u}_n^{(*)} - \hat{\mathbf{u}}_n^{(*)}\big|\big|_{L_2}}{\big|\big|\mathbf{u}_n^{(*)}\big|\big|_{L_2}} \Bigg).
\end{equation}
The training is performed on an NVIDIA V100 (Volta) GPU from the Livermore Computing Lassen system at the Lawrence Livermore National Laboratory, with 3,168 NVIDIA CUDA Cores and 64 GB GDDR5 GPU Memory. 
The open-source TensorFlow library \cite{abadi2016tensorflow} and the Adam optimizer \cite{kingma2014adam} are employed. 
WgLaSDI testing and high-fidelity simulations are performed on an IBM Power9 CPU with 128 cores and 3.5 GHz.

\subsection{1D Burgers' equation}\label{sec:1Dburger}
\begin{figure}[htp]
\centering
    \begin{subfigure}{0.495\textwidth}
        \centering
        \includegraphics[width=1\linewidth]{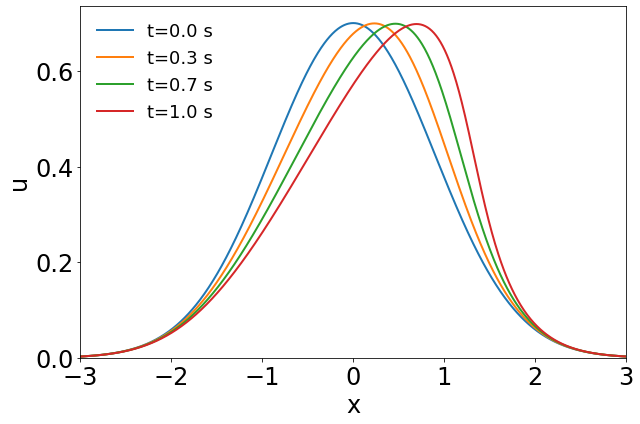}
        \caption{noisy-free}
    \end{subfigure}
    \begin{subfigure}{0.495\textwidth}
        \centering
        \includegraphics[width=1\linewidth]{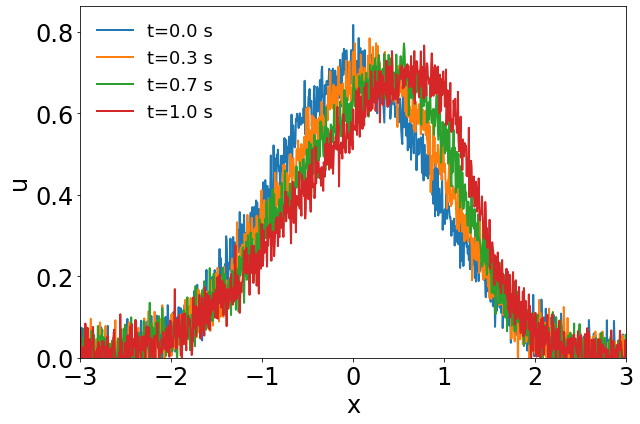}
        \caption{noisy}
    \end{subfigure}
\caption{Physical dynamics of the 1D Burgers' problem with $a=0.7$ and $w=0.9$: (a) noise-free solution; (b) noisy data used for training.}\label{fig.1Dburger_snapshot}
\end{figure}

We first consider a 1D inviscid Burgers' equation with a periodic boundary condition and an initial condition  parameterized by $\boldsymbol{\mu} = \{a, w\} \in \mathcal{D}$:
\begin{equation}\label{eq.1d_burger}
    \begin{cases}
        \displaystyle\frac{\partial u}{\partial t} + u \frac{\partial u}{\partial x} = 0, \quad t \in [0,1], \quad x \in [-3,3] \\
        \displaystyle u(t, x=3) = u(t, x=-3) \\
        \displaystyle u(t=0, x;\boldsymbol{\mu}) = a \exp \bigg(-\frac{x^2}{2w^2}\bigg)
    \end{cases}
\end{equation}
The parameter space in this example is defined as $\mathcal{D}=[0.7,0.9]\times[0.9,1.1]$, which is discretized by $\Delta a=\Delta w=0.01$, leading to a $21 \times 21$ grid of parameters.
High-fidelity simulations adopt a uniform spatial discretization with a nodal spacing of $\Delta x= 6 \times 10^{-3}$ and Backward Euler time integration with $\Delta t= 10^{-3}$. 
A $10\%$ Gaussian white noise component ($\mathcal{N}(0,0.1\sqrt{\sum_n \mathbf{u}_n^2/(N_t+1)})$) is added to the high-fidelity simulation data.
The solution at several time steps of the parameter case $(a=0.7, w=0.9)$ and the corresponding noisy data used for training are shown in Fig. \ref{fig.1Dburger_snapshot}.

\subsubsection{Effects of Weak-Form Dynamics Identification}\label{sec:1Dburger_glasdi_vs_wglasdi}
In the first test, we demonstrate the effects of the weak form on prediction accuracy when dealing with noisy training data. 
Both gLaSDI and WgLaSDI are trained by 16 predefined training samples uniformly distributed on the parameter space $\mathcal{D}=[0.7,0.9]\times[0.9,1.1]$. 
We employ an architecture of 1,001-100-5 ($N_z=5$) with ReLU activation for the encoder and a symmetric architecture for the decoder. 
Linear polynomials are adopted as basis functions for latent space dynamics identification.
The hyperparameters in the WgLaSDI loss function (Eq. \eqref{eq.weak_total_loss}) are defined as $\beta_1 = 1, \beta_2 = 1$, and $\beta_3 = 10^{-5}$, while those in the gLaSDI loss function (Eq. \eqref{eq.total_loss}) are defined as $\beta_1 = 1$ and $\beta_2 = 1$.
We investigated two WgLaSDI loss functions, i.e., \textbf{\textit{Type I}} based on Eqs. \eqref{eq.weak_zdot2} and \eqref{eq.weak_uhatdot} and \textbf{\textit{Type II}} based on Eqs. \eqref{eq.weak_zdot3} and \eqref{eq.weak_uhatdot2}.
The training is performed for 20,000 epochs.
For $k$-NN interpolation of latent space dynamics, $k=4$ is employed for both gLaSDI and WgLaSDI.

\begin{figure}[htp]
\centering
    \begin{subfigure}{0.495\textwidth}
        \centering
        \includegraphics[width=1\linewidth]{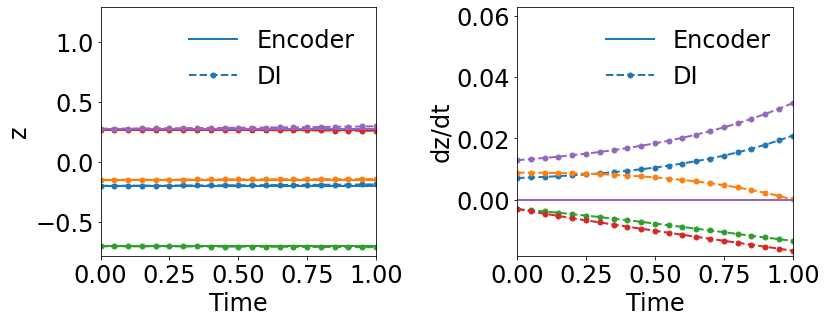}
        \caption{gLaSDI}
    \end{subfigure}
    \begin{subfigure}{0.495\textwidth}
        \centering
        \includegraphics[width=1\linewidth]{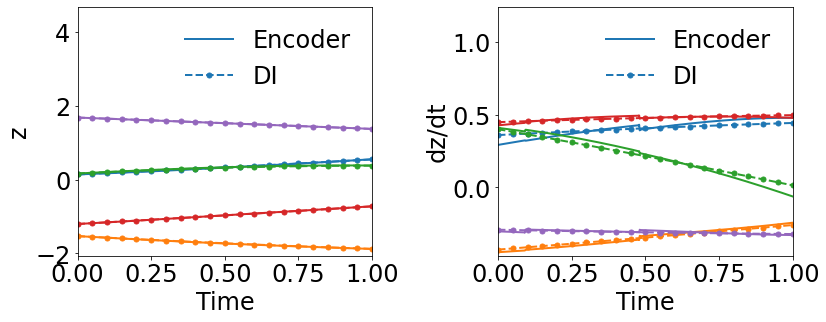}
        \caption{WgLaSDI (Type-I loss)}
    \end{subfigure}
    \begin{subfigure}{0.504\textwidth}
        \centering
        \includegraphics[width=1\linewidth]{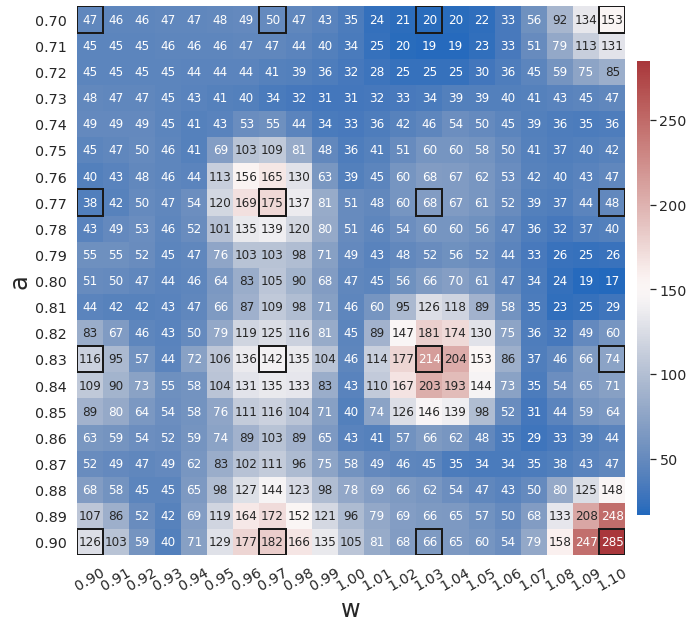}
        \caption{gLaSDI}
    \end{subfigure}
    \begin{subfigure}{0.486\textwidth}
        \centering
        \includegraphics[width=1\linewidth]{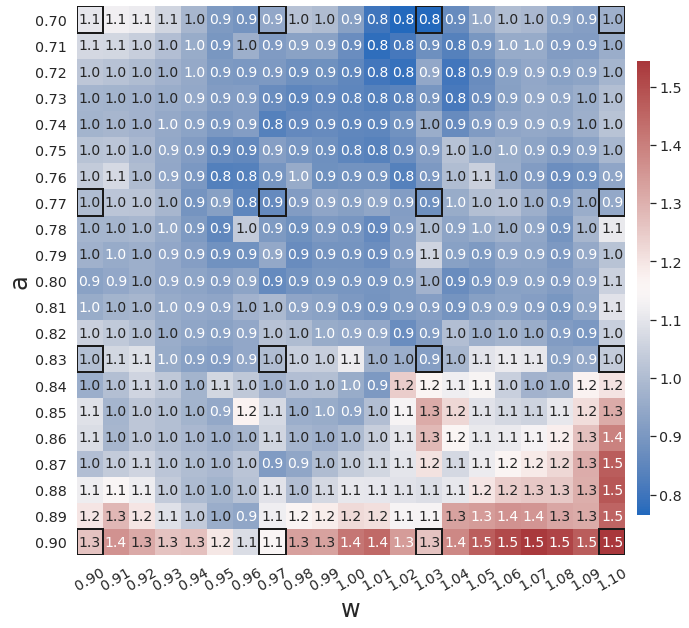}
        \caption{WgLaSDI (Type-I loss)}
    \end{subfigure}
\caption{1D Burgers' problem - The latent space dynamics $(a=0.7, w=0.9)$ predicted by the trained encoder and the DI model from (a) gLaSDI and (b) WgLaSDI based on the Type-I loss function. The maximum relative errors in the parameter space from (c) gLaSDI and (d) WgLaSDI. The black squares indicate the training parameter points.}\label{fig.1Dburger_case1}
\end{figure}

\begin{figure}[htp]
    \centering
    \includegraphics[width=0.5\linewidth]{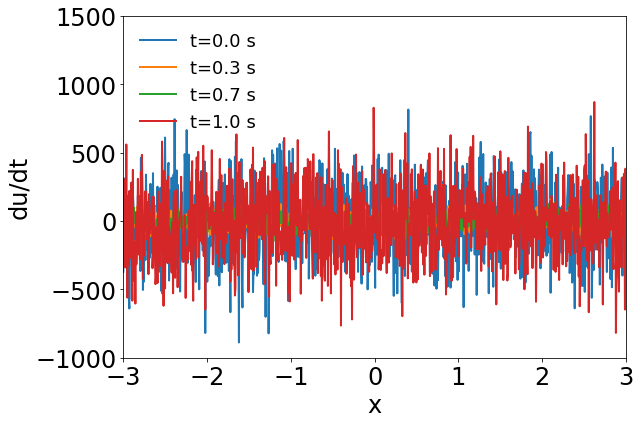}
    \caption{Gradient of noisy dynamics data of the 1D Burgers' problem with $a=0.7$ and $w=0.9$.}
    \label{fig.1Dburger_dyn_grad}
\end{figure}

Fig. \ref{fig.1Dburger_case1} compares the latent space dynamics $(a=0.7, w=0.9)$ and maximum relative errors from gLaSDI and WgLaSDI (based on the Type-I loss function) over the parameter space.
Fig. \ref{fig.1Dburger_case1}(a-b) shows that the DI model (SINDy) in gLaSDI struggles to learn the latent space dynamics predicted by the encoder, while the DI model (WENDy) in WgLaSDI successfully identifies the latent space dynamics.
The consistency between the encoder-predicted and the DI-predicted latent space dynamics is critical to the prediction accuracy of the physical dynamics as reflected in Fig. \ref{fig.1Dburger_case1}(c-d).
WgLaSDI outperforms gLaSDI in the whole parameter space, achieving at most 1.5$\%$ error vs. 285$\%$ of gLaSDI.
It is worth mentioning that the loss function of gLaSDI, as defined in Eq. \eqref{eq.loss_udot} and \eqref{eq.total_loss}, requires the gradient of the dynamics data, which is very oscillatory when the data contains noise, as shown in Fig. \ref{fig.1Dburger_dyn_grad}, leading to difficulties in training.
In contrast, the weak-form loss function of WgLaSDI, as defined in Eq. \eqref{eq.weak_loss_zdot}, \eqref{eq.weak_loss_udot}, and \eqref{eq.weak_total_loss}, avoids the need for the data gradient and takes advantage of the gradient of the test function, which is noise invariant, leading to significantly improved learning performance.
The results of this test demonstrate that the weak form enhances the accuracy and robustness of latent space dynamics identification in the presence of noise.

\begin{figure}[htp]
\centering
    \begin{subfigure}{0.495\textwidth}
        \centering
        \includegraphics[width=1\linewidth]{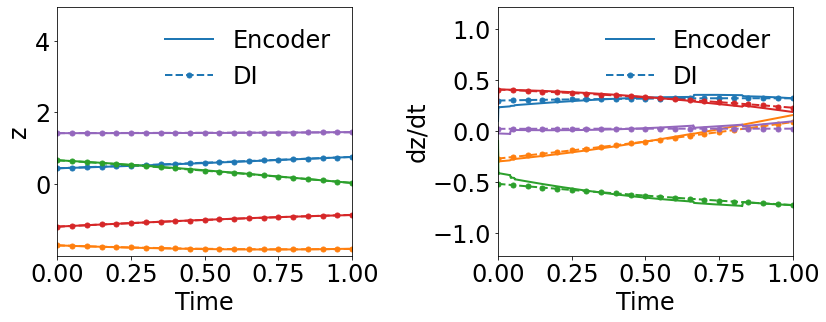}
        \caption{WgLaSDI (Type-II loss)}
    \end{subfigure}
    \begin{subfigure}{0.495\textwidth}
        \centering
        \includegraphics[width=1\linewidth]{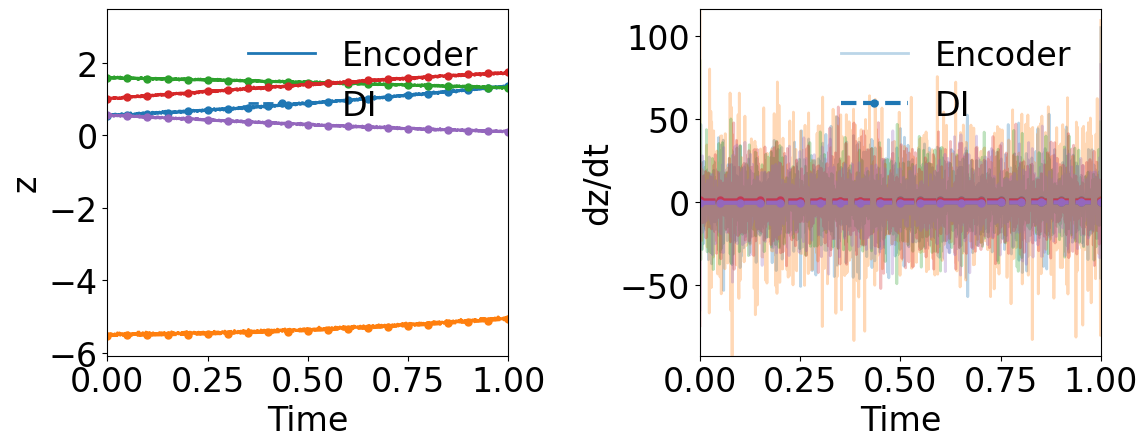}
        \caption{WLaSDI}
    \end{subfigure}
    \begin{subfigure}{0.495\textwidth}
        \centering
        \includegraphics[width=1\linewidth]{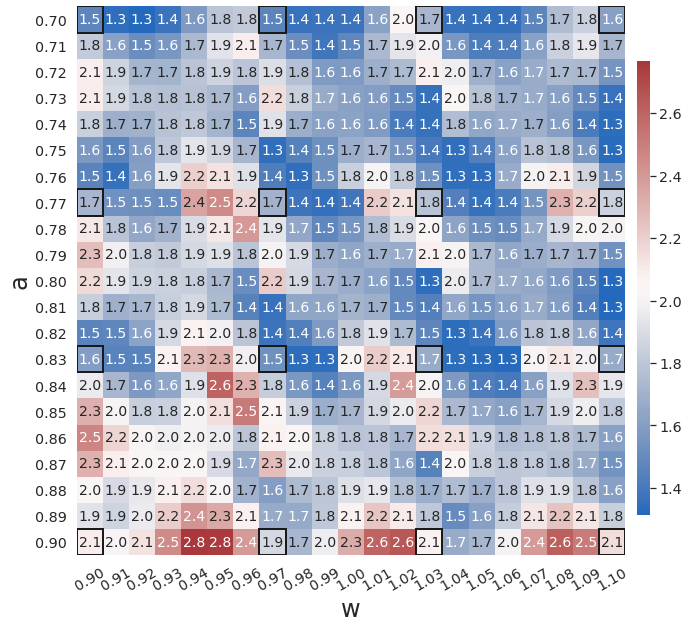}
        \caption{WgLaSDI (Type-II loss)}
    \end{subfigure}
    \begin{subfigure}{0.495\textwidth}
        \centering
        \includegraphics[width=1\linewidth]{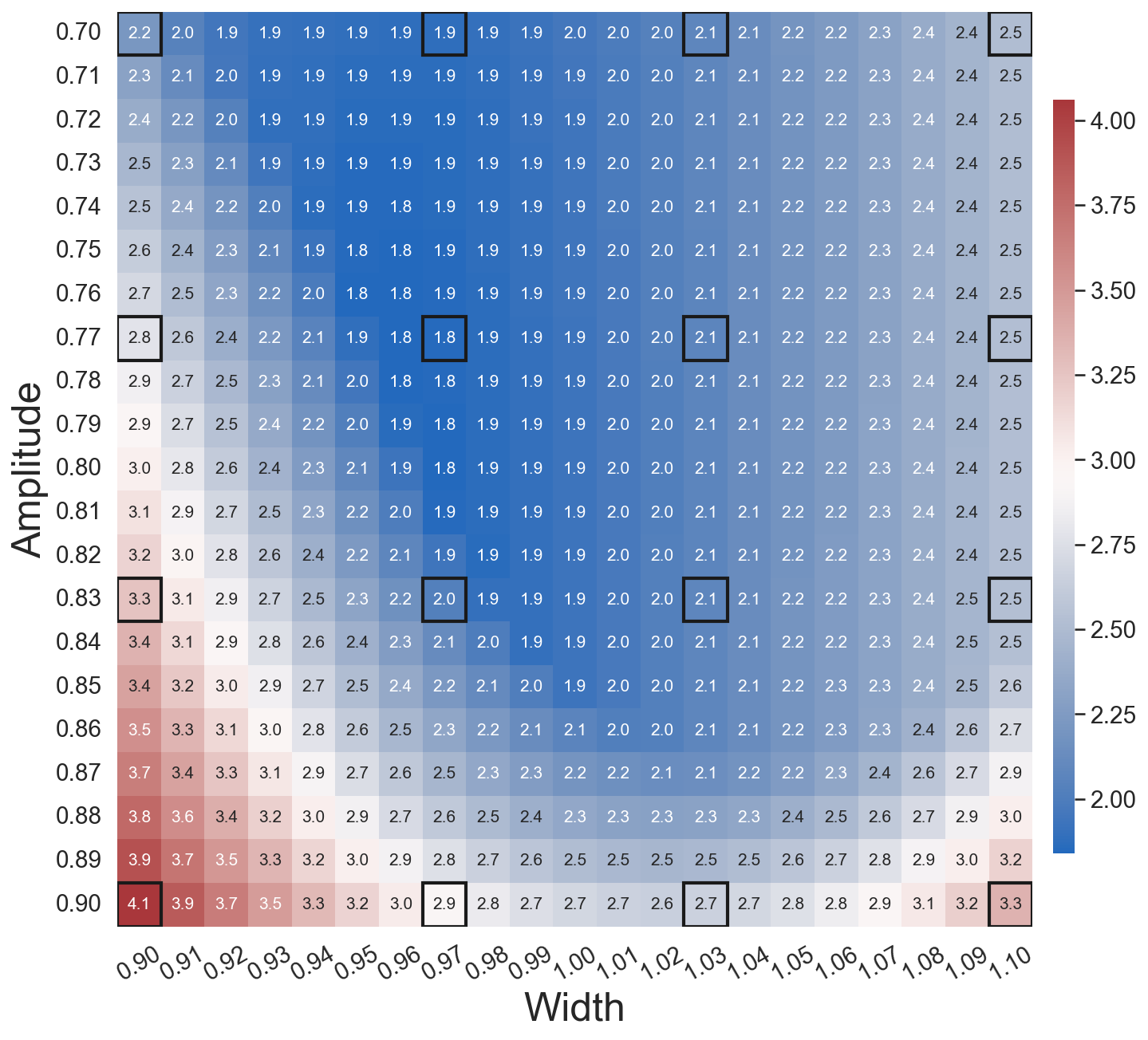}
        \caption{WLaSDI}
    \end{subfigure}
\caption{1D Burgers' problem - The latent space dynamics $(a=0.7, w=0.9)$ predicted by the trained encoder and the DI model from (a) WgLaSDI based on the Type-II loss function and (b) WLaSDI trained on a predefined uniform grid. The maximum relative errors in the parameter space from (c) WgLaSDI and (d) WLaSDI. The black squares indicate the training parameter points.}\label{fig.1Dburger_case1-2}
\end{figure}

Fig. \ref{fig.1Dburger_case1-2}(a) and (c) show that WgLaSDI trained by the Type-II loss function also successfully identifies simple latent space dynamics and achieved a maximum error of 2.8$\%$, slightly higher than that (1.5$\%$) of WgLaSDI based on the Type-I loss function (Fig. \ref{fig.1Dburger_case1}(d)).
The enhanced prediction accuracy achieved by the Type-I loss function is contributed by the involvement of the gradients of the encoder and the decoder networks, leading to stronger constraints on the autoencoder.

\subsubsection{Effects of Simultaneous Training}\label{sec:1Dburger_wlasdi_vs_wglasdi}
In the second test, we compare the effects of fitting the surrogate DI model simultaneously while training the autoencoder versus a sequential training approach. 
Specifically, we compare the weak-form gLaSDI with the weak-form LaSDI \cite{wlasdi}. 
For a fair comparison, hyperparameters such as the autoencoder architecture and the DI basis function, as well as the data are kept the same between the two methods. Fig. \ref{fig.1Dburger_case1-2}(b) and (d) show the results of WLaSDI, with a maximum error of $ 4\%$ across the entire parameter space, slightly larger than those from WgLaSDI for both types of losses. 
The error from WLaSDI includes both the projection error introduced by the autoencoder and the DI error from WENDy, with the projection error being the predominant factor. 
As discussed in Section \ref{sec:1Dburger_glasdi_vs_wglasdi}, it is critical to ensure the consistency between the encoder-predicted and the DI-predicted latent space dynamics for satisfactory prediction accuracy of the physical dynamics.
Due to the lack of interaction between the autoencoder and WENDy of WLaSDI, there is no constraint and regularization from the DI model on the autoencoder, leading to oscillatory latent dynamics from the encoder, as shown in the latent dynamics gradient ($dz/dt$) in Fig. \ref{fig.1Dburger_case1-2}(b).
Although WENDy's variance reduction capability enables it to learn the oscillatory latent dynamics from the encoder, the inconsistency between their predicted latent dynamics still leads to large projection errors of the physical dynamics.
This highlights the impact of using simultaneous training versus the sequential approach. 

\begin{figure}[htp]
\centering
    \begin{subfigure}{0.486\textwidth}
        \centering
        \includegraphics[width=1\linewidth]{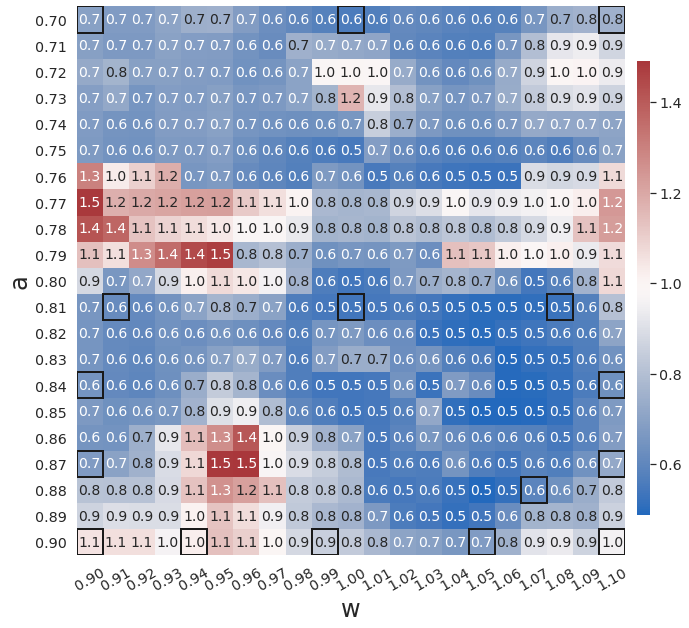}
        \caption{WgLaSDI}
    \end{subfigure}
\caption{1D Burgers' problem - The maximum relative errors in the parameter space from WgLaSDI with physics-informed active learning, based on the Type-I loss function.}\label{fig.1Dburger_case3}
\end{figure}

\subsubsection{Effects of Greedy Physics-Informed Active Learning} \label{sec:1Dburger_active_learning}
In the previous tests, the training samples are predefined on a uniform grid in the parameter space, which may not be optimal, especially for highly nonlinear physical systems.
The greedy physics-informed active learning strategy in WgLaSDI allows maximum exploration of the parameter space and optimal samples to be selected on the fly during the training process. 
To demonstrate that, we train WgLaSDI using the same autoencoder and DI basis functions as defined in Section \ref{sec:1Dburger_glasdi_vs_wglasdi} together with active learning.

WgLaSDI without active learning already achieves excellent performance, with at most 1.5$\%$ maximum relative error over the parameter space, as shown in Fig. \ref{fig.1Dburger_case1}(d). 
WgLaSDI with active learning achieves the same maximum error (1.5$\%$) over the parameter space, as shown in Fig. \ref{fig.1Dburger_case3}, with an average error of 0.75$\%$, slightly lower than that (1.0$\%$) of WgLaSDI without active learning.
It is observed that active learning tends to guide WgLaSDI to select more samples in the higher range of $a$.
The test results show that physics-informed active learning could potentially guide WgLaSDI to select samples for enhanced prediction accuracy over the parameter space.
Compared with the high-fidelity simulation (in-house Python code) that has an around $2\%$ maximum relative error with respect to the high-fidelity data used for WgLaSDI training, WgLaSDI achieves 121$\times$ speed-up.

\subsection{2D Burgers' equation}\label{sec:2Dburger}
\begin{figure}[htp]
\centering
    \begin{subfigure}{0.243\textwidth}
        \centering
        \includegraphics[width=0.78\linewidth]{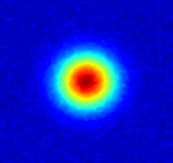}
        \caption{$t=0.0$ s}
    \end{subfigure}
    \begin{subfigure}{0.243\textwidth}
        \centering
        \includegraphics[width=0.78\linewidth]{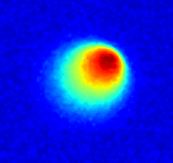}
        \caption{$t=0.2$ s}
    \end{subfigure}
    \begin{subfigure}{0.243\textwidth}
        \centering
        \includegraphics[width=0.78\linewidth]{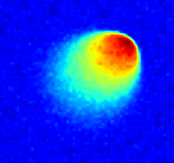}
        \caption{$t=0.8$ s}
    \end{subfigure}
    \begin{subfigure}{0.243\textwidth}
        \centering
        \includegraphics[width=1.0\linewidth]{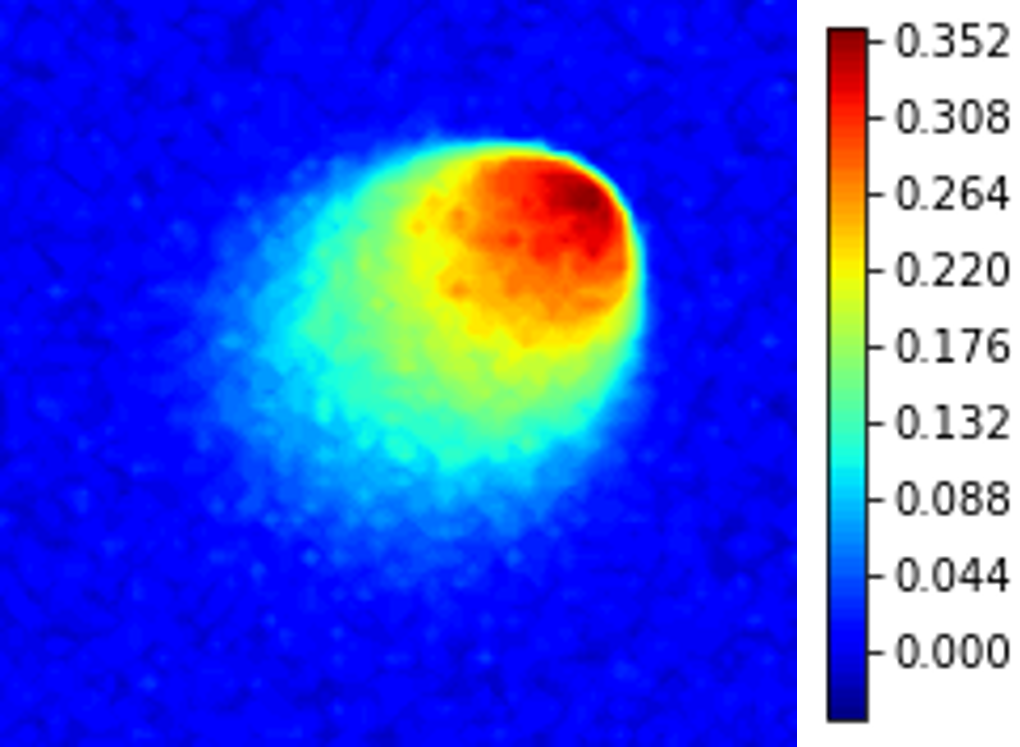}
        \caption{$t=1.0$ s}
    \end{subfigure}
\caption{Noisy physical dynamics of the first velocity component of the 2D Burgers' problem with $a=0.7$ and $w=0.9$: (a) $t=0.0$ s, (b) $t=0.2$ s, (c) $t=0.8$ s, (d) $t=1.0$ s.}\label{fig.2Dburger_snapshot}
\end{figure}

In the second example, we consider a 2D viscous Burgers' equation with an essential boundary condition and an initial condition parameterized by $\boldsymbol{\mu} = \{a, w\} \in \mathcal{D}$:
\begin{equation}\label{eq.2d_burger}
    \begin{cases}
        \displaystyle \frac{\partial \mathbf{u}}{\partial t} + \mathbf{u} \cdot \nabla \mathbf{u} = \frac{1}{Re} \Delta \mathbf{u}, \quad t \in [0,1], \quad \Omega = [-3,3]\times[-3,3] \\
        \displaystyle \mathbf{u}(t, \mathbf{x};\boldsymbol{\mu}) = \mathbf{0} \quad \text{on} \quad \partial\Omega\\
        \displaystyle \mathbf{u}(t=0, \mathbf{x};\boldsymbol{\mu}) = a \exp\bigg({-\frac{||\mathbf{x}||^2}{w^2}}\bigg),
    \end{cases}
\end{equation}
where a Reynolds number of $Re=10,000$ is considered. 
The parameter space in this example is defined as $\mathcal{D}=[0.7,0.9]\times[0.9,1.1]$, which is discretized by $\Delta a=\Delta w=0.01$, leading to a $21 \times 21$ grid of parameters.
The first-order spatial derivative and the diffusion term are approximated by the backward difference and the central difference schemes, respectively.
High-fidelity simulations adopt a uniform spatial discretization with nodal spacing of $\Delta x = \Delta y = 0.1$ and Backward Euler time integration with $\Delta t= 5 \times 10^{-3}$. 
A $10\%$ Gaussian white noise component is added to the high-fidelity simulation data.
The noisy dynamics data of the first velocity component of the parameter case $(a=0.7, w=0.9)$ are shown in Fig. \ref{fig.2Dburger_snapshot}.

\subsubsection{Effects of Weak-Form Dynamics Identification}\label{sec:2Dburger_glasdi_vs_wglasdi}
In the first test, we demonstrate the effects of the weak form on prediction accuracy when dealing with noisy training data. 
Both gLaSDI and WgLaSDI are trained by 25 predefined training samples uniformly distributed on the parameter space $\mathcal{D}=[0.7,0.9]\times[0.9,1.1]$. 
We employ an architecture of 7200-100-5 ($N_z=5$) with ReLU activation for the encoder and a symmetric architecture for the decoder. 
Quadratic polynomials are adopted as basis functions for latent space dynamics identification.
The hyperparameters in the loss functions (Eq. \eqref{eq.weak_total_loss}) are defined as $\beta_1 = 1, \beta_2 = 1$, and $\beta_3 = 10^{-5}$, while those in the gLaSDI loss function (Eq. \eqref{eq.total_loss}) are defined as $\beta_1 = 0$ and $\beta_2 = 1$.
We investigated two WgLaSDI loss functions, i.e., \textit{Type-I} based on Eqs. \eqref{eq.weak_zdot2} and \eqref{eq.weak_uhatdot} and \textit{Type-II} based on Eqs. \eqref{eq.weak_zdot3} and \eqref{eq.weak_uhatdot2}.
The training is performed for 100,000 epochs.
For $k$-NN interpolation of latent space dynamics, $k=4$ is employed for both gLaSDI and WgLaSDI.

\begin{figure}[htp]
\centering
    \begin{subfigure}{0.495\textwidth}
        \centering
        \includegraphics[width=1\linewidth]{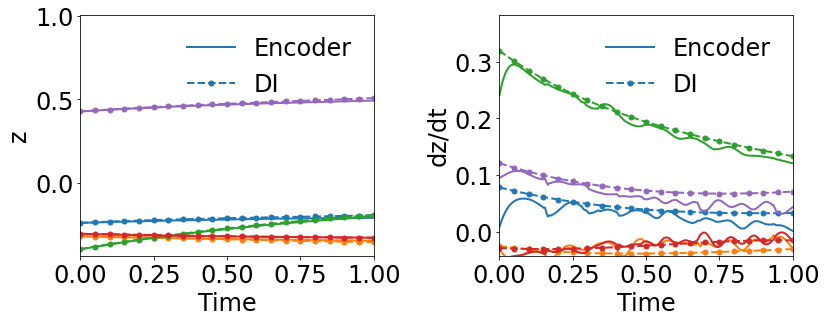}
        \caption{gLaSDI}
    \end{subfigure}
    \begin{subfigure}{0.495\textwidth}
        \centering
        \includegraphics[width=1\linewidth]{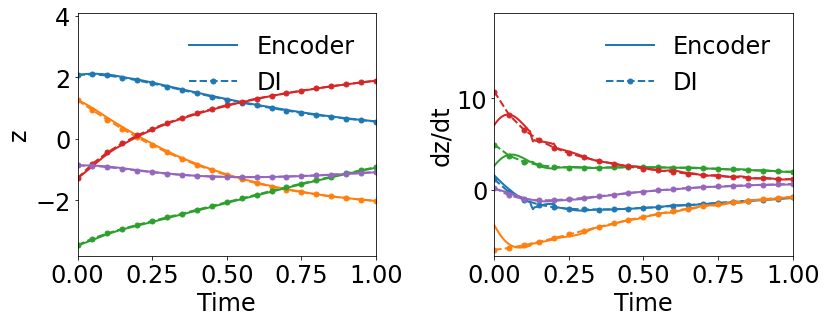}
        \caption{WgLaSDI (Type-I loss)}
    \end{subfigure}
    \begin{subfigure}{0.504\textwidth}
        \centering
        \includegraphics[width=1\linewidth]{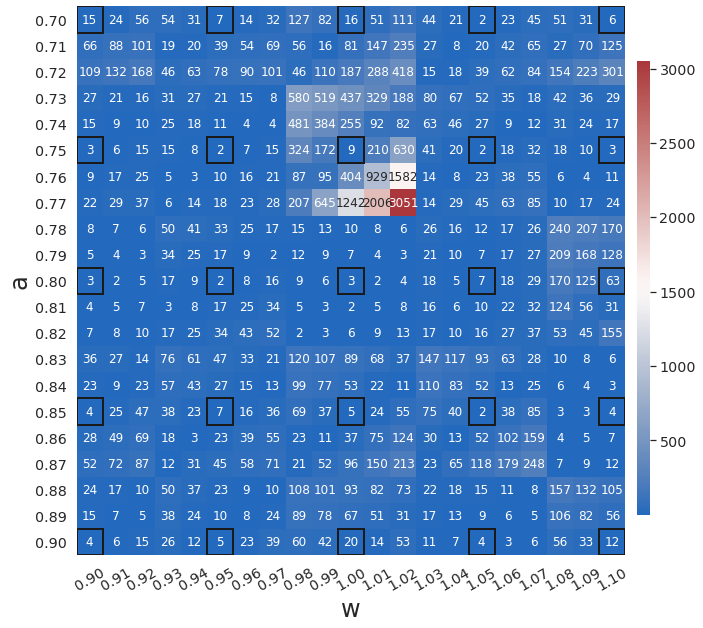}
        \caption{gLaSDI}
    \end{subfigure}
    \begin{subfigure}{0.486\textwidth}
        \centering
        \includegraphics[width=1\linewidth]{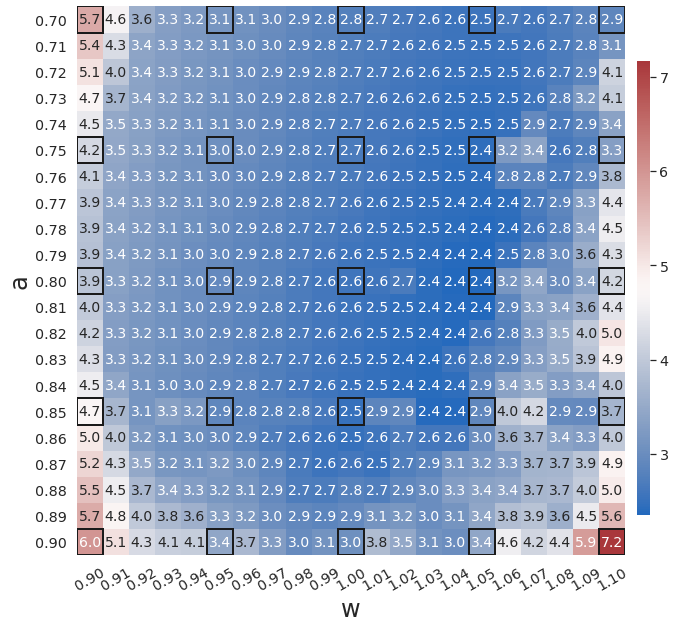}
        \caption{WgLaSDI (Type-I loss)}
    \end{subfigure}
\caption{2D Burgers' problem - The latent space dynamics $(a=0.7, w=0.9)$ predicted by the trained encoder and the DI model from (a) gLaSDI and (b) WgLaSDI based on the Type-I loss function. The maximum relative errors in the parameter space from (c) gLaSDI and (d) WgLaSDI. The black squares indicate the training parameter points.}\label{fig.2Dburger_case1}
\end{figure}

Fig. \ref{fig.2Dburger_case1} shows that WgLaSDI (based on the Type-I loss function) outperforms gLaSDI in learning latent space dynamics in the presence of noise, achieving at most 7.2$\%$ error vs. 3,051$\%$ of gLaSDI across the parameter space.
It further demonstrates the enhanced robustness and performance against noise due to the weak-form latent space dynamics identification.

\begin{figure}[htp]
\centering
    \begin{subfigure}{0.495\textwidth}
        \centering
        \includegraphics[width=1\linewidth]{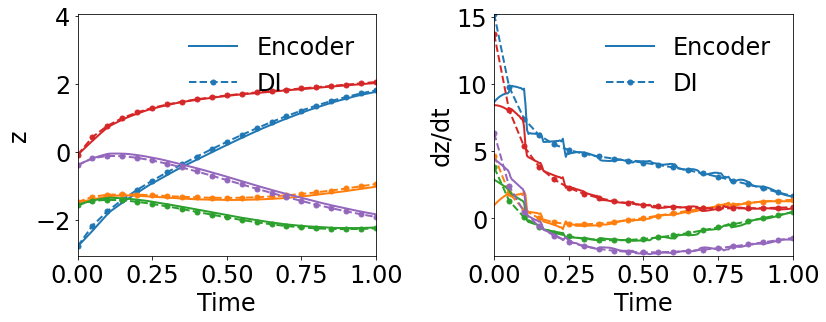}
        \caption{WgLaSDI (Type-II loss)}
    \end{subfigure}
    \begin{subfigure}{0.495\textwidth}
        \centering
        \includegraphics[width=1\linewidth]{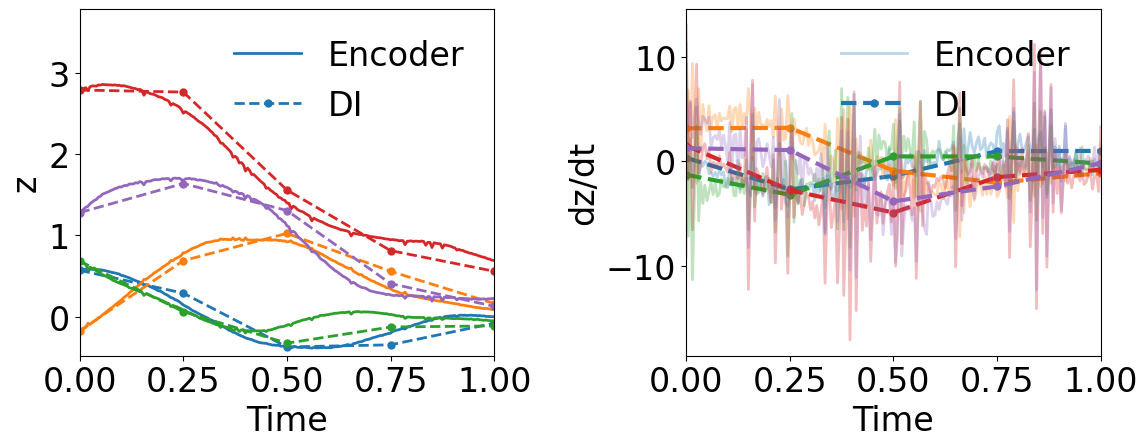}
        \caption{WLaSDI}
    \end{subfigure}
    \begin{subfigure}{0.495\textwidth}
        \centering
        \includegraphics[width=1\linewidth]{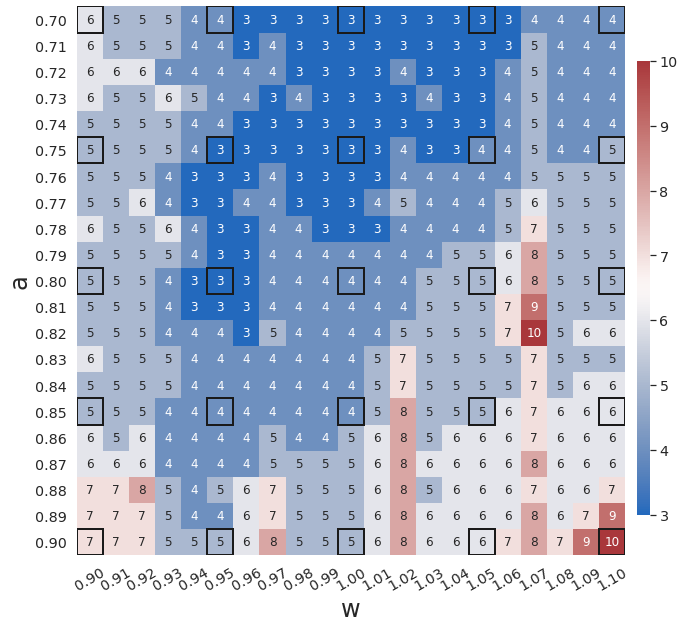}
        \caption{WgLaSDI (Type-II loss)}
    \end{subfigure}
    \begin{subfigure}{0.495\textwidth}
        \centering
        \includegraphics[width=1\linewidth]{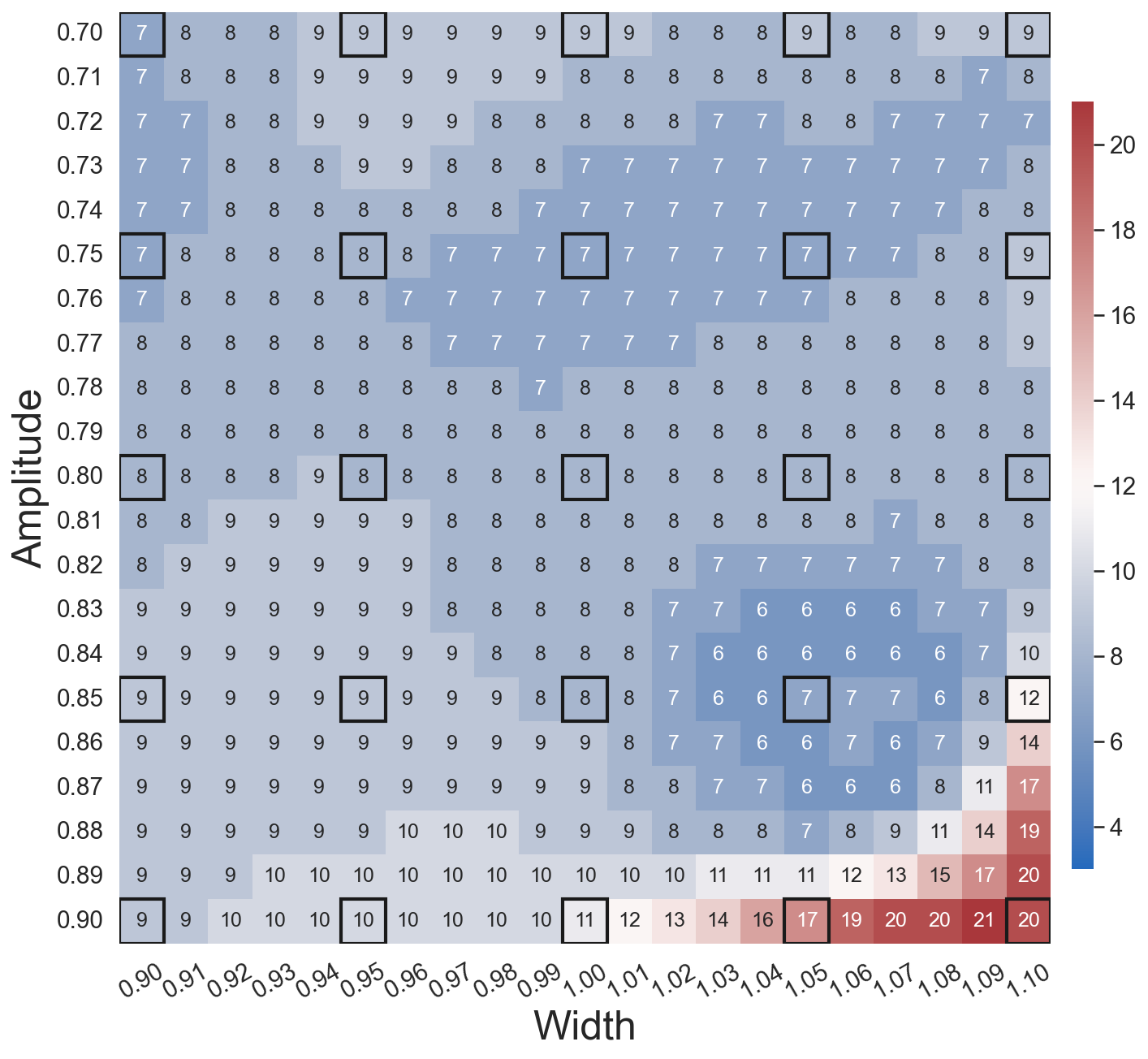}
        \caption{WLaSDI}
    \end{subfigure}
\caption{2D Burgers' problem - The latent space dynamics $(a=0.7, w=0.9)$ predicted by the trained encoder and the DI model from (a) WgLaSDI based on the Type-II loss function and (b) WLaSDI trained on a predefined uniform grid. The maximum relative errors in the parameter space from (c) WgLaSDI and (d) WLaSDI. The black squares indicate the training parameter points.}\label{fig.2Dburger_case1-2}
\end{figure}

Fig. \ref{fig.2Dburger_case1-2}(a) shows the latent space dynamics identified by WgLaSDI based on the Type-II loss function, which is more oscillatory than that from the Type-I loss (Fig. \ref{fig.2Dburger_case1}(b)).
Fig. \ref{fig.2Dburger_case1-2}(c) shows that WgLaSDI based on the Type-II loss achieves a maximum error of 10$\%$, higher than that (7.2$\%$) from the Type-I loss (Fig. \ref{fig.2Dburger_case1}(d)).
It further demonstrates the enhanced performance achieved by the Type-I loss function.
We will adopt the Type-I loss function for WgLaSDI in the following examples.

\subsubsection{Effects of Simultaneous Training}\label{sec:2Dburger_wlasdi_vs_wglasdi}
In the second test, we further demonstrate the effects of simultaneous training of the autoencoder and the DI model by comparing WgLaSDI with WLaSDI. 
WLaSDI is trained using the same autoencoder, DI basis functions, and data as those employed for WgLaSDI in Section \ref{sec:2Dburger_glasdi_vs_wglasdi}. 
As shown in Fig. \ref{fig.2Dburger_case1-2}(d), the WLaSDI error ranges from 6$\%$ to 20$\%$, significantly larger than that of WgLaSDI for both types of losses (Fig. \ref{fig.2Dburger_case1}(d) and Fig. \ref{fig.2Dburger_case1-2}(c)). 
Comparing Fig. \ref{fig.2Dburger_case1}(b) and Fig. \ref{fig.2Dburger_case1-2}(a) with Fig. \ref{fig.2Dburger_case1-2}(b) shows that simultaneous training enforces constraints on the latent space dynamics discovered by the autoencoder through the interaction with the DI model, enabling simpler latent space dynamics, reduced projection and DI errors, and therefore better generalization performance. 

\begin{figure}[htp]
\centering
    \begin{subfigure}{0.486\textwidth}
        \centering
        \includegraphics[width=1\linewidth]{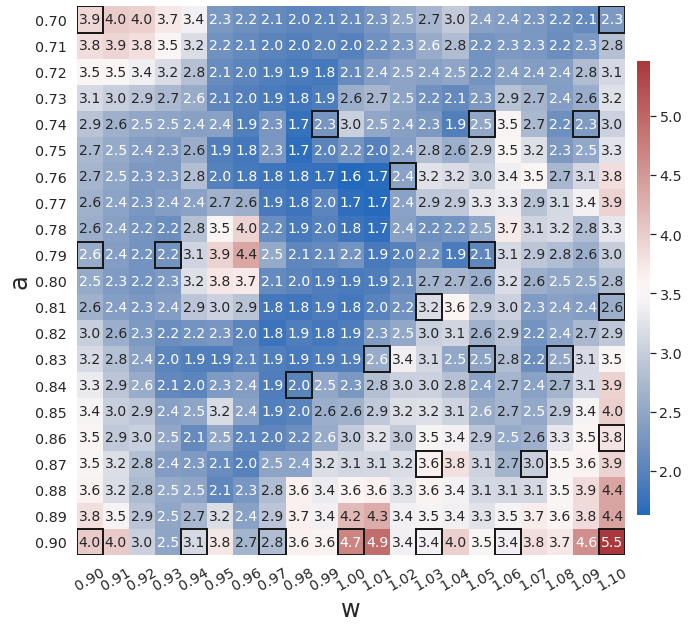}
        \caption{WgLaSDI}
    \end{subfigure}
\caption{2D Burgers' problem - The maximum relative errors in the parameter space from WgLaSDI with greedy physics-informed active learning, based on the Type-I loss function.}\label{fig.2Dburger_case3}
\end{figure}

\subsubsection{Effects of Greedy Physics-Informed Active Learning} \label{sec:2Dburger_active_learning}
To demonstrate the effects of the greedy physics-informed active learning strategy in WgLaSDI on the model performance, we train WgLaSDI using the same autoencoder and DI basis functions defined in Section \ref{sec:2Dburger_glasdi_vs_wglasdi} together with active learning. 
Fig. \ref{fig.2Dburger_case3} shows that WgLaSDI with active learning achieves at most 5.5$\%$ error over the parameter space, lower than that (7.2$\%$) of WgLaSDI without active learning, as shown in Fig. \ref{fig.2Dburger_case1}(d).
Interestingly, active learning tends to guide WgLaSDI to select more samples in the higher range of $w$ to achieve better performance in the parameter space.
Compared with the high-fidelity simulation (in-house Python code) that has an around $6\%$ maximum relative error with respect to the high-fidelity data used for WgLaSDI training, WgLaSDI achieves 1,374$\times$ speed-up.

\subsection{Time-dependent radial advection}\label{sec:advection}
\begin{figure}[htp]
\centering
    \begin{subfigure}{0.243\textwidth}
        \centering
        \includegraphics[width=0.81\linewidth]{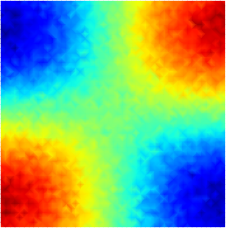}
        \caption{$t=0.0$ s}
    \end{subfigure}
    \begin{subfigure}{0.243\textwidth}
        \centering
        \includegraphics[width=0.81\linewidth]{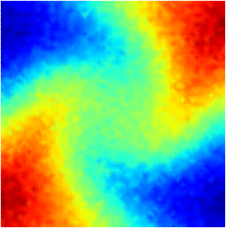}
        \caption{$t=0.7$ s}
    \end{subfigure}
    \begin{subfigure}{0.243\textwidth}
        \centering
        \includegraphics[width=0.81\linewidth]{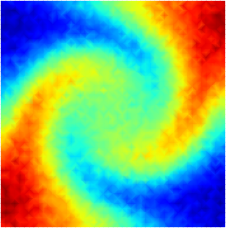}
        \caption{$t=1.5$ s}
    \end{subfigure}
    \begin{subfigure}{0.243\textwidth}
        \centering
        \includegraphics[width=1.0\linewidth]{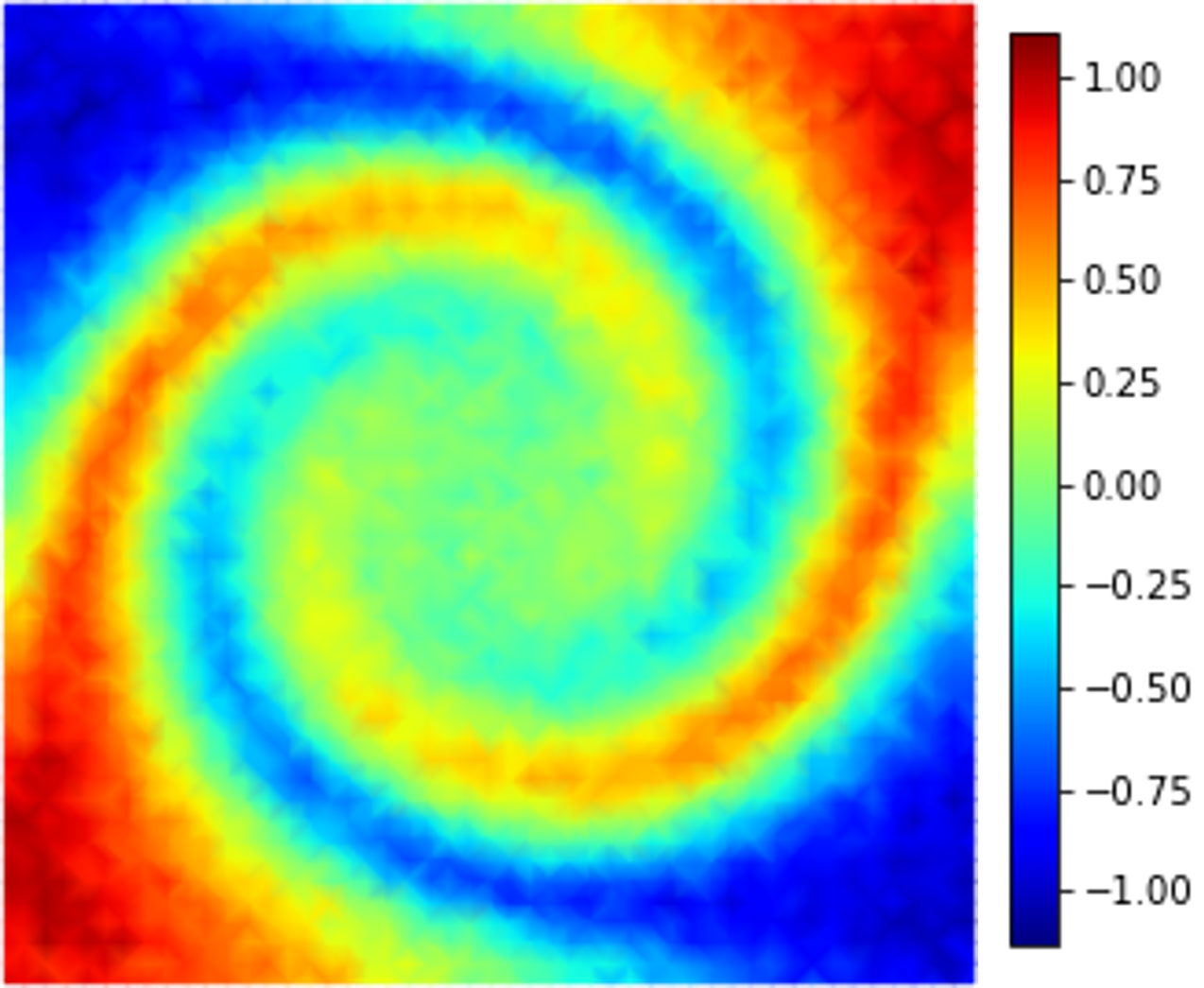}
        \caption{$t=3.0$ s}
    \end{subfigure}
\caption{Noisy physical dynamics of the time-dependent radial advection problem with $w_1=1.5$ and $w_2=2.0$: (a) $t=0.0$ s, (b) $t=0.7$ s, (c) $t=1.5$ s, (d) $t=3.0$ s.}\label{fig.advection_snapshot}
\end{figure}

In the third example, we consider a 2D time-dependent radial advection problem with an initial condition parameterized by $\boldsymbol{\mu} = \{w_1, w_2\} \in \mathcal{D}$:
\begin{equation}\label{eq.advection}
    \begin{cases}
        \displaystyle \frac{\partial u}{\partial t} + \mathbf{v} \cdot \nabla  u = 0, \quad t \in [0,3], \quad \Omega = [-1,1]\times[-1,1] \\
        \displaystyle u(t, \mathbf{x};\boldsymbol{\mu}) = 0 \quad \text{on} \quad \partial\Omega \\
        u(t=0, \mathbf{x};\boldsymbol{\mu}) = \sin(w_1 x_1) \sin(w_2 x_2),
    \end{cases}
\end{equation}
where $\mathbf{v} = \frac{\pi}{2} d [x_2, -x_1]^T$ denotes the fluid velocity and $d = (1-x_1^2)^2(1-x_2^2)^2$.
The parameter space in this example is defined as $\mathcal{D}=[1.5,2.0]\times[2.0,2.5]$, which is discretized by $\Delta w_1 = \Delta w_2 = 0.025$, leading to a $21 \times 21$ grid of parameters.
In high-fidelity simulations, the spatial domain is discretized by first-order periodic square finite elements constructed on a uniform grid of $96 \times 96$ discrete points. 
The fourth-order Runge-Kutta explicit time integrator with a uniform time step of $\Delta t = 0.01$ is employed.
A $10\%$ Gaussian white noise component is added to the high-fidelity simulation data.
The noisy dynamics data of the parameter case $(w_1=1.5, w_2=2.0)$ are shown in Fig. \ref{fig.advection_snapshot}.
 
\subsubsection{Effects of Weak-Form Dynamics Identification}\label{sec:advection_glasdi_vs_wglasdi}
To examine the effects of the weak form on model performance when dealing with noisy data, we train gLaSDI and WgLaSDI using 16 predefined samples uniformly distributed on the parameter space $\mathcal{D}=[1.5,2.0]\times[2.0,2.5]$. 
Both models adopt an architecture of 9,216-100-3 ($N_z=3$) with ReLU activation for the encoder and a symmetric architecture for the decoder, with linear polynomials as basis functions for latent space dynamics identification.
The hyperparameters in the WgLaSDI loss function (Eq. \eqref{eq.weak_total_loss}) are defined as $\beta_1 = 1, \beta_2 = 1$, and $\beta_3 = 10^{-4}$, while those in the gLaSDI loss function (Eq. \eqref{eq.total_loss}) are defined as $\beta_1 = 1$ and $\beta_2 = 1$.
The training is performed for 100,000 epochs.
For $k$-NN interpolation of latent space dynamics, $k=4$ is employed for both gLaSDI and WgLaSDI.

\begin{figure}[htp]
\centering
    \begin{subfigure}{0.495\textwidth}
        \centering
        \includegraphics[width=1\linewidth]{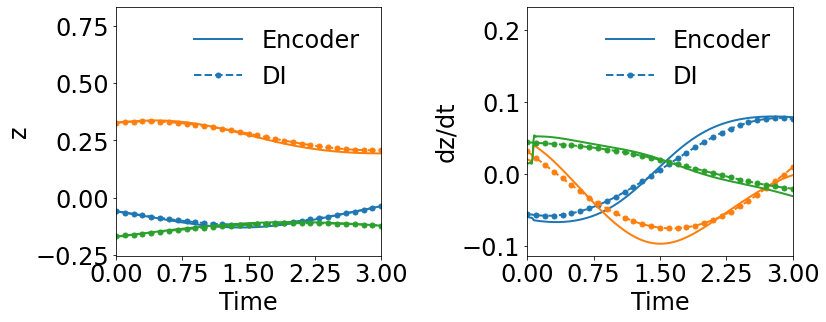}
        \caption{gLaSDI}
    \end{subfigure}
    \begin{subfigure}{0.495\textwidth}
        \centering
        \includegraphics[width=1\linewidth]{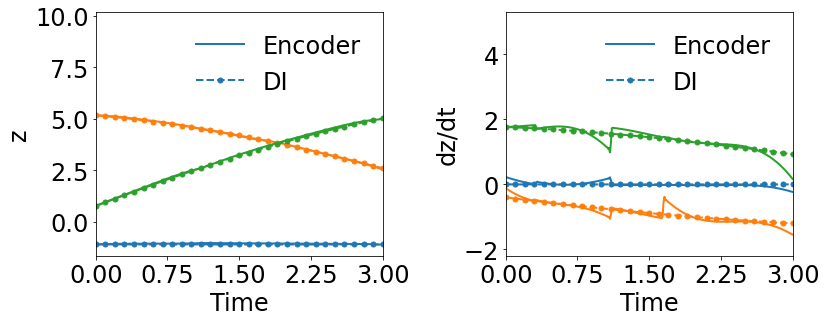}
        \caption{WgLaSDI}
    \end{subfigure}
    \begin{subfigure}{0.495\textwidth}
        \centering
        \includegraphics[width=1\linewidth]{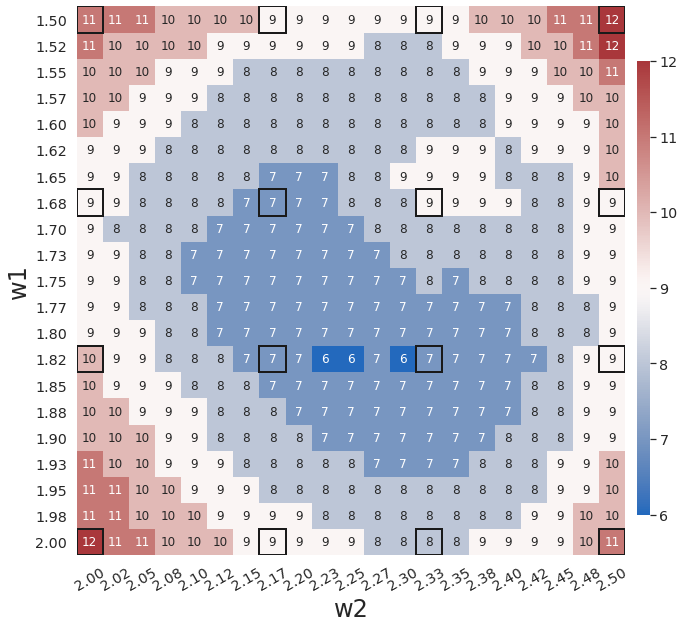}
        \caption{gLaSDI}
    \end{subfigure}
    \begin{subfigure}{0.495\textwidth}
        \centering
        \includegraphics[width=1\linewidth]{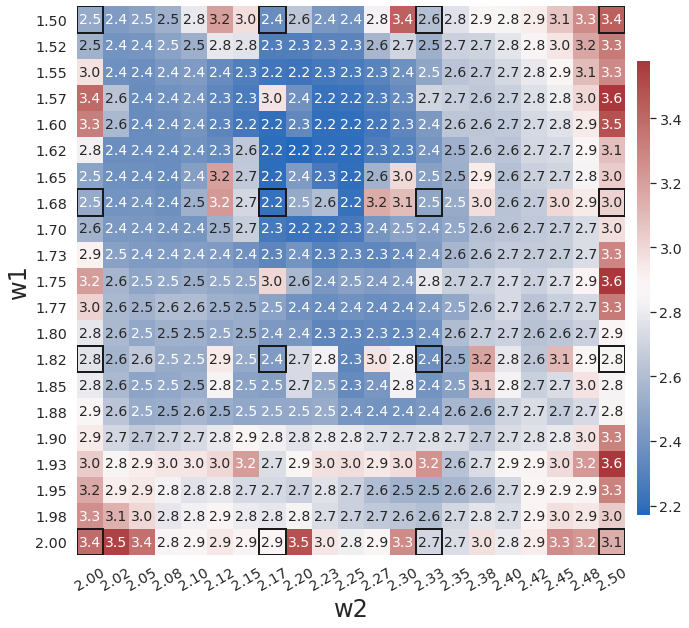}
        \caption{WgLaSDI}
    \end{subfigure}
\caption{Time-dependent radial advection problem - The latent space dynamics $(w_1=1.5, w_2=2.0)$ predicted by the trained encoder and the DI model from (a) gLaSDI and (b) WgLaSDI. The maximum relative errors in the parameter space from (c) gLaSDI and (d) WgLaSDI. The black squares indicate the training parameter points.}\label{fig.advection_case1}
\end{figure}

Fig. \ref{fig.advection_case1} shows that WgLaSDI outperforms gLaSDI in learning latent space dynamics in the presence of noise, achieving at most 3.6$\%$ error vs. 12.6$\%$ of gLaSDI across the parameter space.
WgLaSDI demonstrates enhanced robustness and performance against noise due to the weak-form dynamics identification.

\subsubsection{Effects of Simultaneous Training}\label{sec:advection_wlasdi_vs_wglasdi}
\begin{figure}[htp]
\centering
    \begin{subfigure}{0.495\textwidth}
        \centering
        \includegraphics[width=1\linewidth]{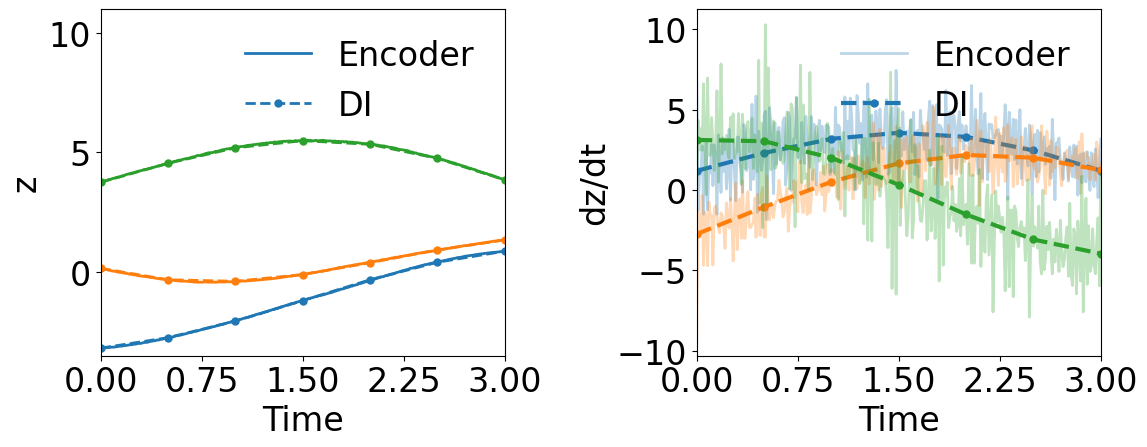}
        \caption{gLaSDI}
    \end{subfigure}
    \begin{subfigure}{0.495\textwidth}
        \centering
        \includegraphics[width=1\linewidth]{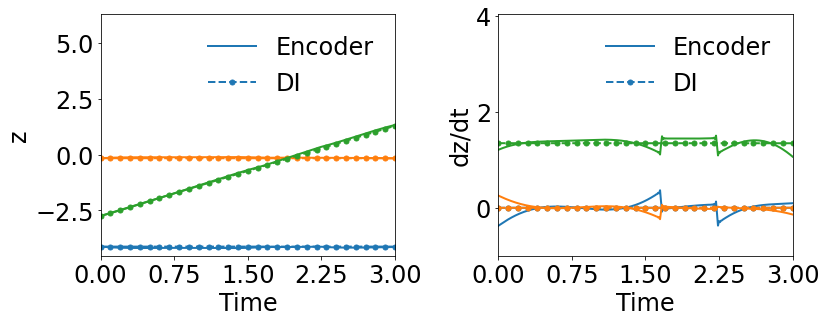}
        \caption{WgLaSDI}
    \end{subfigure}
    \begin{subfigure}{0.495\textwidth}
        \centering
        \includegraphics[width=1\linewidth]{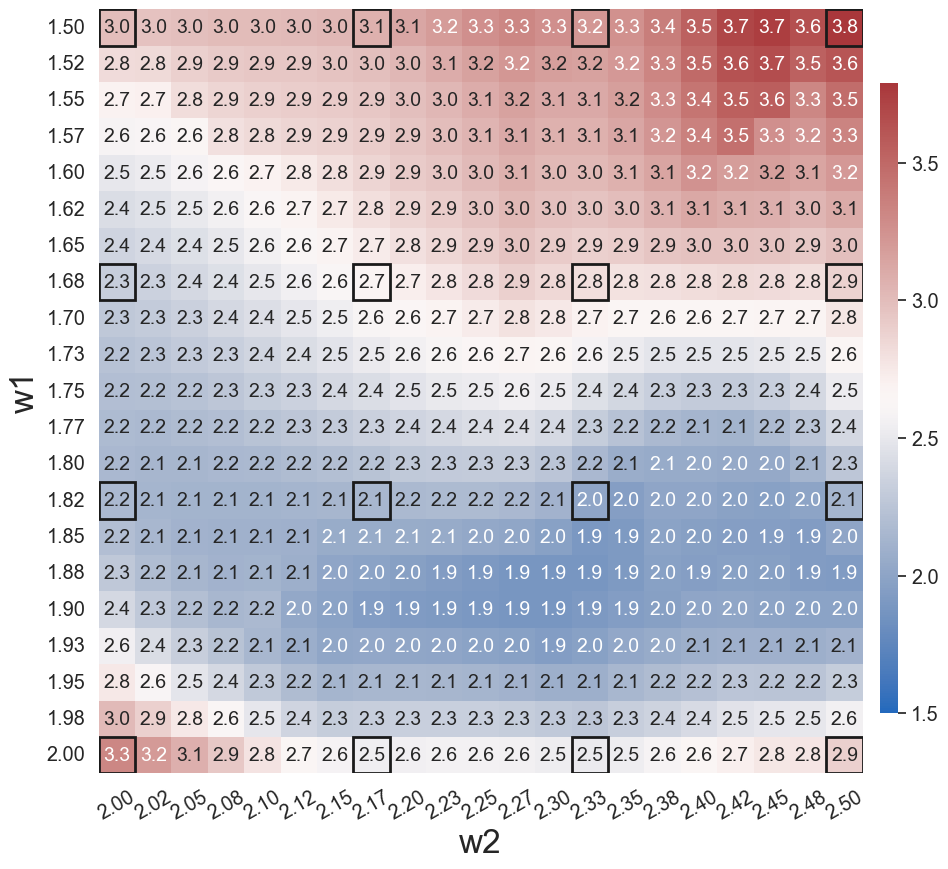}
        \caption{WLaSDI}
    \end{subfigure}
    \begin{subfigure}{0.495\textwidth}
        \centering
        \includegraphics[width=1\linewidth]{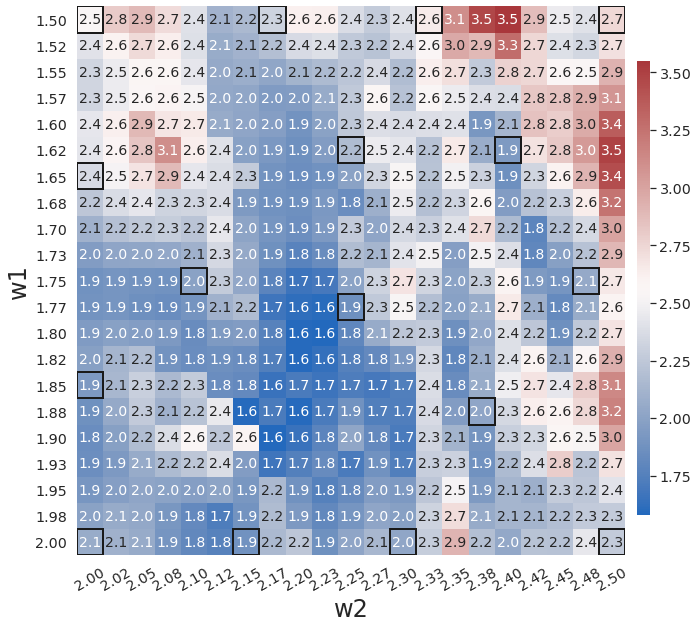}
        \caption{WgLaSDI}
    \end{subfigure}
\caption{Time-dependent radial advection problem - The latent space dynamics $(w_1=1.5, w_2=2.0)$ predicted by the trained encoder and the DI model from (a) WLaSDI trained on a predefined uniform grid and (b) WgLaSDI with greedy physics-informed active learning. The maximum relative errors in the parameter space from (c) WLaSDI and (d) WgLaSDI. The black squares indicate the training parameter points.}\label{fig.advection_case2-3}
\end{figure}

The effect of simultaneous training is further illustrated in this example of radial advection. 
Both WgLaSDI and WLaSDI employ the same autoencoder, DI basis functions, and data described in Section \ref{sec:advection_glasdi_vs_wglasdi}.
According to Fig. \ref{fig.advection_case2-3}(c), WLaSDI achieves a maximum error of 3.8$\%$ across the entire parameter space, slightly higher than that (3.6$\%$) of WgLaSDI (Fig. \ref{fig.advection_case1}(d)).
It further demonstrates the effectiveness and advantages of simultaneous training.

\subsubsection{Effects of Greedy Physics-Informed Active Learning} \label{sec:advection_active_learning}
To further demonstrate the effects of the greedy physics-informed active learning on model performance, we train WgLaSDI using the same autoencoder and DI basis functions defined in Section \ref{sec:2Dburger_glasdi_vs_wglasdi} together with active learning.
Fig. \ref{fig.advection_case2-3} shows that physics-informed active learning helps WgLaSDI to further reduce the maximum error from 3.6$\%$ to 3.5$\%$ and the average error from 2.7$\%$ to 2.2$\%$ in the parameter space.
Compared with the high-fidelity simulation (MFEM \cite{anderson2021mfem}) that has an around $3\%$ maximum relative error with respect to the high-fidelity data used for WgLaSDI training, the WgLaSDI model achieves 178$\times$ speed-up.

\subsection{1D-1V Vlasov Equation}\label{sec:vlasov}
\begin{figure}[htp]
\centering
    \begin{subfigure}{0.243\textwidth}
        \centering
        \includegraphics[width=0.81\linewidth]{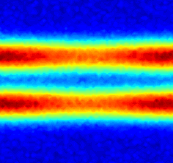}
        \caption{$t=0.0$ s}
    \end{subfigure}
    \begin{subfigure}{0.243\textwidth}
        \centering
        \includegraphics[width=0.81\linewidth]{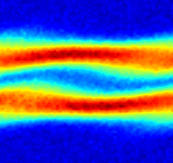}
        \caption{$t=1.2$ s}
    \end{subfigure}
    \begin{subfigure}{0.243\textwidth}
        \centering
        \includegraphics[width=0.81\linewidth]{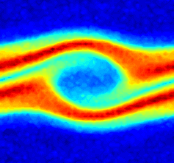}
        \caption{$t=3.7$ s}
    \end{subfigure}
    \begin{subfigure}{0.243\textwidth}
        \centering
        \includegraphics[width=1.0\linewidth]{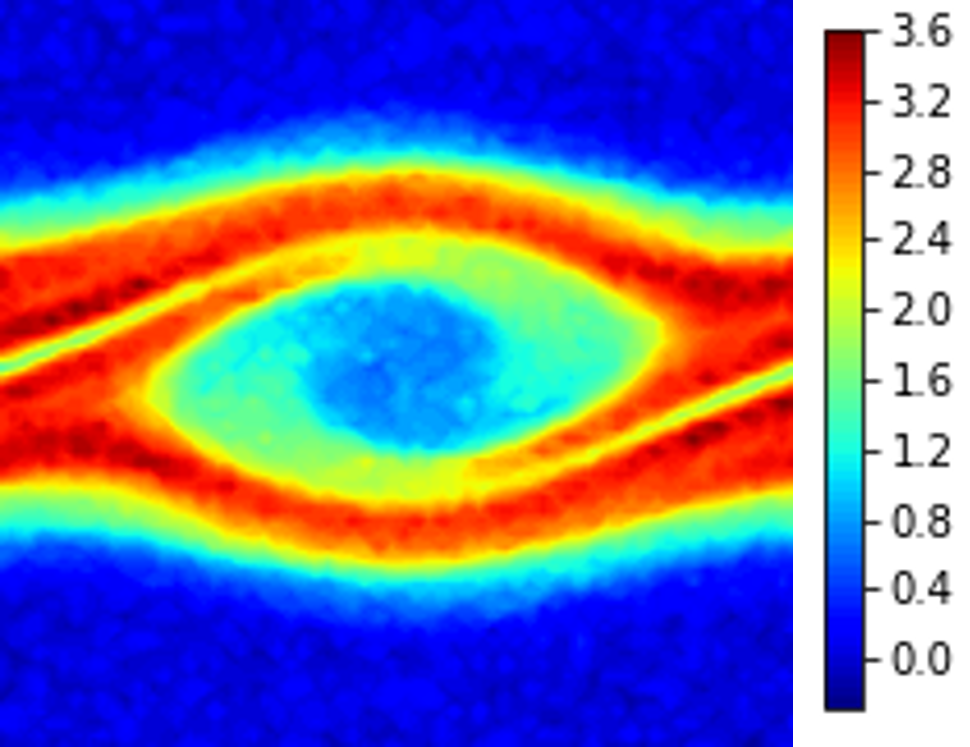}
        \caption{$t=5.0$ s}
    \end{subfigure}
\caption{Noisy physical dynamics of the 1D-1V Vlasov-Poisson equation with $T=0.9$ and $k=1.0$: (a) $t=0.0$ s, (b) $t=1.2$ s, (c) $t=3.7$ s, (d) $t=5.0$ s.}\label{fig.vlasov_snapshot}
\end{figure}

In the last example, we consider a simplified 1D-1V Vlasov-Poisson equation with an initial condition parameterized by $\boldsymbol{\mu} = \{T, k\} \in \mathcal{D}$:
\begin{equation}\label{eq.vlasov}
    \begin{cases}
        \displaystyle \frac{\partial f}{\partial t} + \frac{\partial vf}{\partial x} + \frac{\partial}{\partial v} \bigg(\frac{d \Phi}{dx}f\bigg) = 0, \quad t \in [0,5], \quad x \in [0,2\pi] \quad v \in [-7,7] \\
        \displaystyle \frac{d^2 \Phi}{d x^2} = \int_v f dv \\
        \displaystyle f(t=0, x, v;\boldsymbol{\mu}) = \frac{8}{\sqrt{2 \pi T}} \bigg[ 1 + \frac{1}{10} \cos(k x) \bigg] \bigg[ \exp \bigg( -\frac{(v-2)^2}{2T} \bigg) +
        \exp \bigg( -\frac{(v+2)^2}{2T} \bigg) \bigg],
    \end{cases}
\end{equation}
where $f(x,v)$ is the plasma distribution function, dependent on a spatial coordinate $x$ and a velocity coordinate $v$ and $\Phi$ is the electrostatic potential.
This simplified model describes 1D collisionless electrostatic plasma dynamics, which is representative of complex models for plasma behaviors occurring in nuclear fusion reactors.
This is a 2D PDE due to the velocity variable.
The parameter space in this example is defined as $\mathcal{D}=[0.9,1.1]\times[1.0,1.2]$, which is discretized by $\Delta k = \Delta T = 0.01$, leading to a $21 \times 21$ grid of parameters.
High-fidelity simulations are performed using the HyPar solver \cite{HyPar} with a WENO spatial discretization \cite{jiangshu} and the fourth-order Runge-Kutta explicit time integration scheme ($\Delta t = 5 \times 10^{-3}$).
A $5\%$ Gaussian white noise component is added to the high-fidelity simulation data.
The noisy dynamics data of the parameter case $(T=0.9, k=1.0)$ are shown in Fig. \ref{fig.vlasov_snapshot}.

\subsubsection{Effects of Weak-Form Dynamics Identification}\label{sec:vlasov_glasdi_vs_wglasdi}
To examine the effects of the weak form on model performance when dealing with noisy training data, we train gLaSDI and WgLaSDI using 16 predefined samples uniformly distributed on the parameter space $\mathcal{D}=[0.9,1.1]\times[1.0,1.2]$. 
Both models adopt an architecture of 4,096-1000-100-50-50-50-3 ($N_z=3$) with ReLU activation for the encoder and a symmetric architecture for the decoder, with linear polynomials as basis functions for latent space dynamics identification.
The hyperparameters in the WgLaSDI loss function (Eq. \eqref{eq.weak_total_loss}) are defined as $\beta_1 = 10, \beta_2 = 100$, and $\beta_3 = 10^{-5}$, while those in the gLaSDI loss function (Eq. \eqref{eq.total_loss}) are defined as $\beta_1 = 0$ and $\beta_2 = 10$.
The training is performed for 400,000 epochs.
For $k$-NN interpolation of latent space dynamics, $k=4$ is employed for both gLaSDI and WgLaSDI.

\begin{figure}[htp]
\centering
    \begin{subfigure}{0.495\textwidth}
        \centering
        \includegraphics[width=1\linewidth]{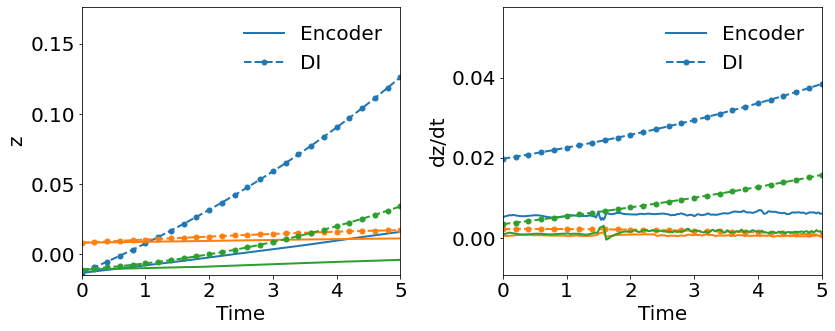}
        \caption{gLaSDI}
    \end{subfigure}
    \begin{subfigure}{0.495\textwidth}
        \centering
        \includegraphics[width=1\linewidth]{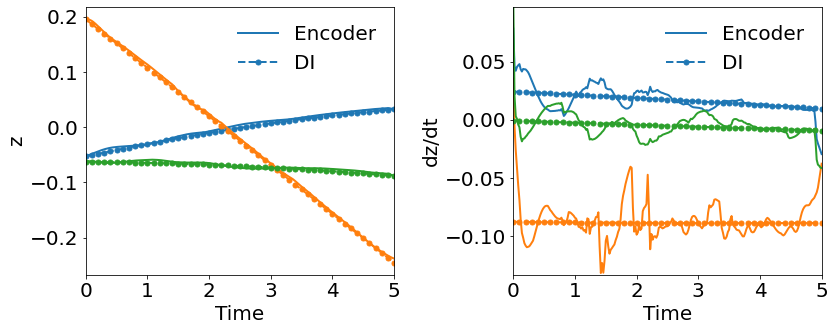}
        \caption{WgLaSDI}
    \end{subfigure}
    \begin{subfigure}{0.495\textwidth}
        \centering
        \includegraphics[width=1\linewidth]{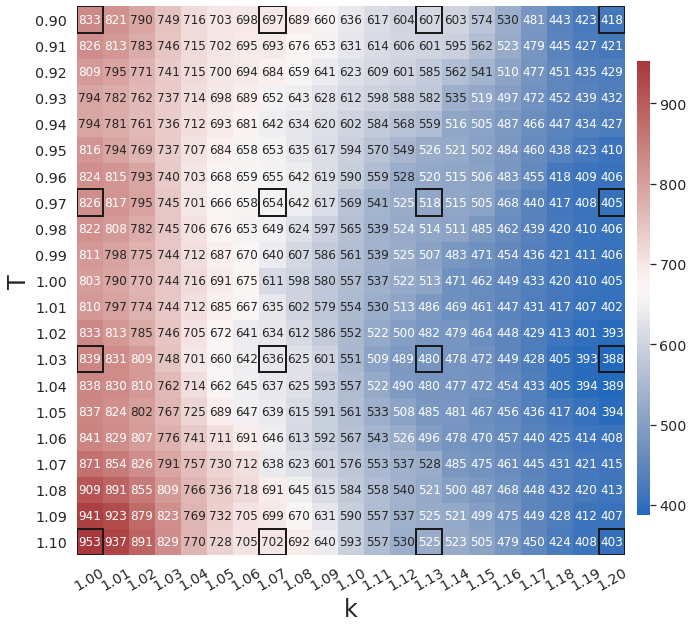}
        \caption{gLaSDI}
    \end{subfigure}
    \begin{subfigure}{0.495\textwidth}
        \centering
        \includegraphics[width=1\linewidth]{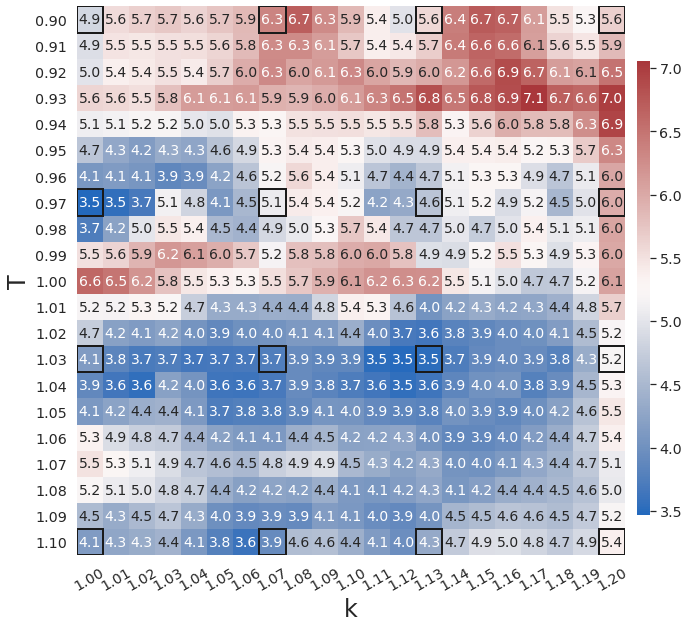}
        \caption{WgLaSDI}
    \end{subfigure}
\caption{1D-1V Vlasov equation - The latent space dynamics $(T=0.9, k=1.0)$ predicted by the trained encoder and the DI model from (a) gLaSDI and (b) WgLaSDI. The maximum relative errors in the parameter space from (c) gLaSDI and (d) WgLaSDI. The black squares indicate the training parameter points.}\label{fig.vlasov_case1}
\end{figure}

Fig. \ref{fig.vlasov_case1} shows that WgLaSDI achieves at most 7.1$\%$ error in the parameter space, significantly lower than 953$\%$ of gLaSDI.
WgLaSDI again demonstrates enhanced robustness and performance against noise.
Compared with the high-fidelity simulation (HyPar \cite{HyPar}), the WgLaSDI model achieves 1,779$\times$ speed-up.

\subsubsection{Effects of Simultaneous Training}\label{sec:vlasov_wlasdi_vs_wglasdi}
\begin{figure}[htp]
\centering
\begin{subfigure}{0.495\textwidth}
     \centering
     \includegraphics[width=1\linewidth]{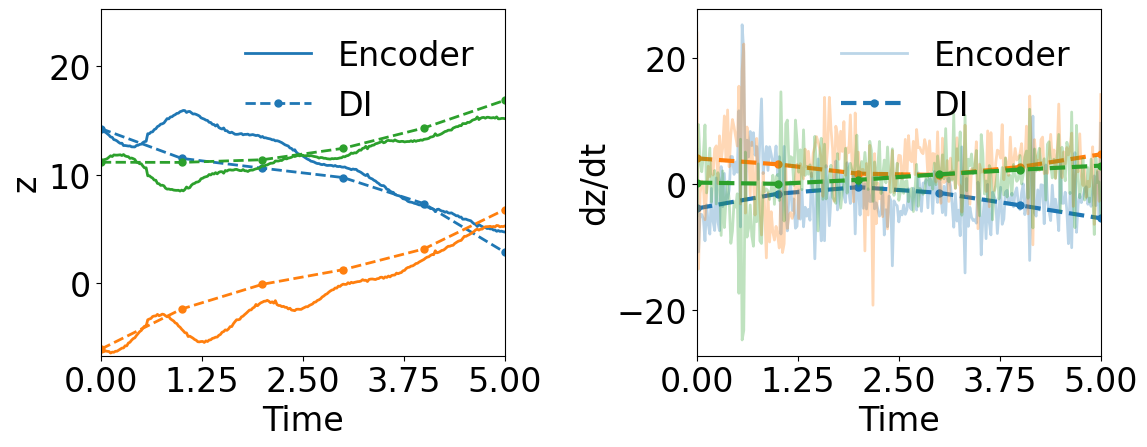}
     \caption{WLaSDI}
 \end{subfigure}
 \begin{subfigure}{0.495\textwidth}
     \centering
     \includegraphics[width=1\linewidth]{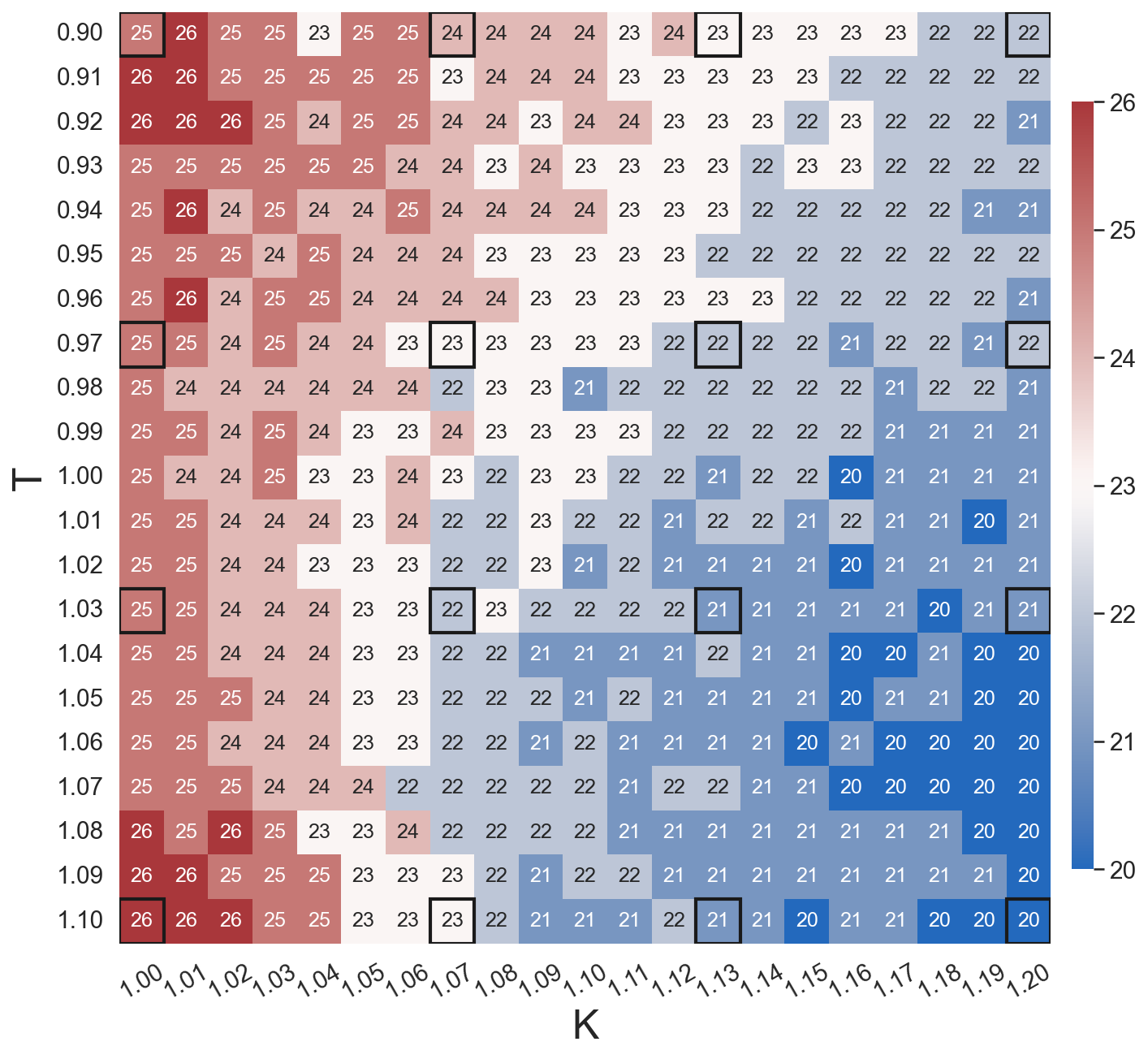}
     \caption{WLaSDI}
 \end{subfigure}
\caption{1D-1V Vlasov equation - (a) The latent space dynamics $(T=0.9, k=1.0)$ predicted by the trained encoder and the DI model from WLaSDI; (b) The maximum relative errors in the parameter space from WLaSDI. The black squares indicate the training parameter points.}\label{fig.vlasov_case2}
\end{figure}

We trained a WLaSDI with the same autoencoder, DI basis functions, and data described in Section \ref{sec:vlasov_glasdi_vs_wglasdi}.
The large discrepancy between the encoder-predicted and the DI-predicted latent space dynamics from WLaSDI, as shown in Fig. \ref{fig.vlasov_case2}(a), leads to larger errors (20$\%$ to 26$\%$) across the parameter space (Fig. \ref{fig.vlasov_case2}(b)) than those (3.5$\%$ - 7.1$\%$) of WgLaSDI (Fig. \ref{fig.vlasov_case1}(d)).
It again highlights that simultaneous training enhances latent space dynamics learning and contributes to simpler latent space dynamics with a stronger connection and consistency between the autoencoder and the DI model, leading to better performance.

\section{Conclusions}\label{sec:conclusion}
In this study, we introduced a weak-form greedy latent-space dynamics identification (WgLaSDI) framework to enhance the accuracy and robustness of data-driven ROM against noisy data. 
Built upon gLaSDI, the proposed WgLaSDI framework consists of an autoencoder for nonlinear dimensionality reduction and WENDy for latent-space dynamics identification.
The autoencoder and WENDy models are trained interactively and simultaneously through minimizing a weak-form loss function defined in Eq. \eqref{eq.weak_total_loss}, enabling identification of simple latent-space dynamics for improved accuracy, efficiency, and robustness of data-driven computing. 
The greedy physics-informed active learning based on a residual-based error indicator facilitates parameter space exploration and searches for the optimal training samples on the fly.

We demonstrated that the proposed WgLaSDI framework is applicable to a wide range of physical phenomena, including viscous and inviscid Burgers' equations, time-dependent radial advection, and two-stream plasma instability.
We investigated the effects of various key ingredients of the WgLaSDI framework, including the weak-form latent space dynamics identification, simultaneous training of the autoencoder and DI models, and greedy physics-informed active learning, by comparing WgLaSDI with gLaSDI and WLaSDI in terms of their performance against noisy data.
WgLaSDI demonstrated superior performance, with enhanced accuracy and robustness, achieving 1-7$\%$ relative errors and 121 to 1,779$\times$ speed-up (to the high-fidelity models). 

Although the parametrization in this study considers only the parameters from the initial conditions, the parameterization can be easily extended to account for other parameterization types, such as governing equations and material properties, as demonstrated in \cite{he2023glasdi}, which could be useful for inverse problems.
To further improve the proposed framework, automatic neural architecture search \cite{elsken2019neural} could be leveraged to optimize the autoencoder architecture automatically and maximize generalization performance.
Although the greedy physics-informed active learning in WgLaSDI offers great advantages for optimal performance with minimal data, it cannot be applied to noisy data when the physics is unknown.
Extension with data-driven active learning based on uncertainty quantification, such as GPLaSDI \cite{bonneville2024gplasdi}, could be a promising solution.

\section*{Acknowledgements}
This work was supported in part by a Rudy Horne Fellowship to AT. This work also received partial support from the U.S. Department of Energy, Office of Science, Office of Advanced Scientific Computing Research, as part of the CHaRMNET Mathematical Multifaceted Integrated Capability Center (MMICC) program, under Award Number DE-SC0023164 to YC at Lawrence Livermore National Laboratory, and under Award Number DE-SC0023346 to DMB at the University of Colorado Boulder. Lawrence Livermore National Laboratory is operated by Lawrence Livermore National Security, LLC, for the U.S. Department of Energy, National Nuclear Security Administration under Contract DE-AC52-07NA27344. IM release number: 
LLNL-JRNL-865859. This work also utilized the Blanca condo computing resource at the University of Colorado Boulder. Blanca is jointly funded by computing users and the University of Colorado Boulder, United States.

\section*{Data Availability}
The dataset of the 1D Burgers' equation and 2D Burgers' equation can be generated using the open-source gLaSDI code available at \url{https://github.com/LLNL/gLaSDI}.
The dataset of the time-dependent radial advection problem can generated using the open-source MFEM code (example 9) available at \url{https://github.com/mfem/mfem/tree/master}.
The dataset of the Vlasov equation for plasma physics can be generated using the open-source HyPar solver available at \url{https://github.com/debog/hypar}.

\section*{Conflict of interest}
The authors declare that they have no conflict of interest.

\appendix

\printbibliography

\end{document}